\documentclass[%
 reprint,
 amsmath,amssymb,
 aps, prl, superscriptaddress
]{revtex4-2}

\usepackage{graphicx}
\usepackage{dcolumn}
\usepackage{bm}

\usepackage{xcolor}
\usepackage{hyperref}
\usepackage{array}

\begin{document}

\preprint{APS/123-QED}

\title{Uncovering Antipolar Ordering and Pressure-Tunable Phases in Hexagonal LaN}

\author{Atanu Paul}
\affiliation{Smart Ferroic Materials Center, Physics Department and Institute for Nanoscience and Engineering, University of Arkansas, Fayetteville, Arkansas 72701, USA}
\author{Laurent Bellaiche}
\affiliation{Smart Ferroic Materials Center, Physics Department and Institute for Nanoscience and Engineering, University of Arkansas, Fayetteville, Arkansas 72701, USA}
\affiliation{Department of Materials Science and Engineering, Tel Aviv University, Ramat Aviv, Tel Aviv 6997801, Israel}
\author{Charles Paillard}
\affiliation{Smart Ferroic Materials Center, Physics Department and Institute for Nanoscience and Engineering, University of Arkansas, Fayetteville, Arkansas 72701, USA}
\affiliation{Université Paris-Saclay, CentraleSupélec, CNRS, Laboratoire SPMS, Gif-sur-Yvette, 91190 France.}

\date{\today}

\begin{abstract}

We predict an antipolar instability in hexagonal LaN using first-principles density functional theory. Starting from a nonpolar hexagonal phase, we identify competing polar and antipolar zone-center phonon instabilities. Condensation of the polar and antipolar modes stabilizes, respectively, dynamically stable wurtzite (WZ) phase and an hexagonal antipolar (AP) phase which is characterized by alternating local polarization and zero net macroscopic polarization within the unit cell.
At ambient conditions, the AP phase is metastable with respect to the WZ phase, and a finite energy barrier exists between these phases, suggesting a possible polarization-switching pathway via the AP intermediate state. The energy barrier between the WZ and AP phases decreases with increasing pressure, indicating enhanced tunability between polar and antipolar states. The sublattice polarization increases with pressure in the AP phase, while it decreases in the WZ phase. We further find that, with increasing pressure, the rock-salt and tetragonal phases of LaN become more stable than the hexagonal phases (AP and WZ). Consequently, the realization of the AP phase is more favorable in the low-pressure regime, where hexagonal phases remain energetically competitive. These results demonstrate pressure-driven competition between polar and antipolar phases in LaN and point toward antiferroelectric-like behavior in this binary nitride system.

\end{abstract}

\maketitle



Group III-V nitrides have attracted immense interest in advanced semiconductor technology owing to their wide and tunable band gaps, high electron mobility, and exceptional chemical and thermal stability~\cite{RevModPhys.87.1139, RevModPhys.87.1119, Liu2026, Thermal_Conduct, Amano_2020, band_Gap_Engi}, making them indispensable for electronic and optoelectronic applications~\cite{ZHAO201514}. In the wurtzite (WZ) structure, these nitrides lack inversion symmetry, exhibiting spontaneous polarization and strong piezoelectricity, sensitive to strain and chemical composition. Although binary nitrides like AlN and GaN are non-switchable under normal conditions, compositional engineering can induce ferroelectricity, as demonstrated in Al\({}_{1-x}\)Sc\({}_{x}\)N~\cite{AlSCN_switch}. This breakthrough bridges wide-band-gap semiconductors and functional ferroelectrics (FEs) for next-generation energy-efficient, non-volatile memory devices ~\cite{FE_Next, FE_Next1}.

Antiferroelectrics (AFEs) offer an alternative paradigm, featuring antiparallel dipole sublattices with zero macroscopic polarization that can transform into a polar state under an electric field~\cite{Catalan2026, AntiFerro, AFE_Rabe}. This field-induced phase transition yields a characteristic double hysteresis loop, making AFEs highly attractive for high-energy-density capacitors and electrocaloric applications~\cite{SI2024101231, AP_review}. However, AFE materials remain rare, traditionally confined to complex oxides like PbZrO$_{3}$, AgNbO$_{3}$, NaNbO$_{3}$ or La$_{3}$NbO$_{7}$~\cite{PZO, PZO_Anti, AgNbO, NaNbO, AFE_2026}. Discovering AFE phases within semiconductor-compatible, binary, or alloyed nitride platforms, remains an important challenge.
Lanthanum nitride (LaN) is a promising candidate within this expanded nitride family, exhibiting potential for high-performance capacitors~\cite{acsenergylett}. While LaN ground-state crystallizes in the rock-salt (RS) structure at ambient conditions~\cite{jacs_1952}, WZ and zinc-blende phases have been experimentally realized ~\cite{Krause_nb5222}. High-pressure studies revealed a structural phase transition from a cubic RS to a tetragonal symmetry~\cite{LaN_press}. First-principles calculations have predicted low-symmetry ($P$1) distortions of the rock-salt phase~\cite{CHEN2021110779} and potential ferroelectricity in the WZ and $P$1 phases~\cite{CHEN2021110779, PhysRevMaterials.5.094602}.

In this work, we use density functional theory (DFT) to demonstrate that the hexagonal nonpolar (H5) phase \cite{Peng} of LaN hosts both zone-centered antipolar and polar lattice instabilities (See Supplemental Material~\cite{SM} for Methods). Condensation of the antipolar mode yields a dynamically stable antipolar (AP) phase. Although the energy of this AP phase lies above the polar WZ ground state within hexagonal symmetry, the calculated energy landscape reveals a finite activation barrier, suggesting accessibility via non-equilibrium growth, chemical doping, or properly designed excitation. Importantly, we show that external hydrostatic pressure tunes the relative stability between the WZ and AP phases, making the AP phase energetically favorable. This represents the first prediction of an antipolar instability in a hexagonal binary nitride, establishing a new pathway toward antiferroelectricity in rare-earth nitride platforms. Another AP phase has also been predicted for LaN under high pressure; however, it occurs in an orthorhombic structure~\cite{Ding2022}.

\begin{figure}[t]
\includegraphics[width=8.3cm]{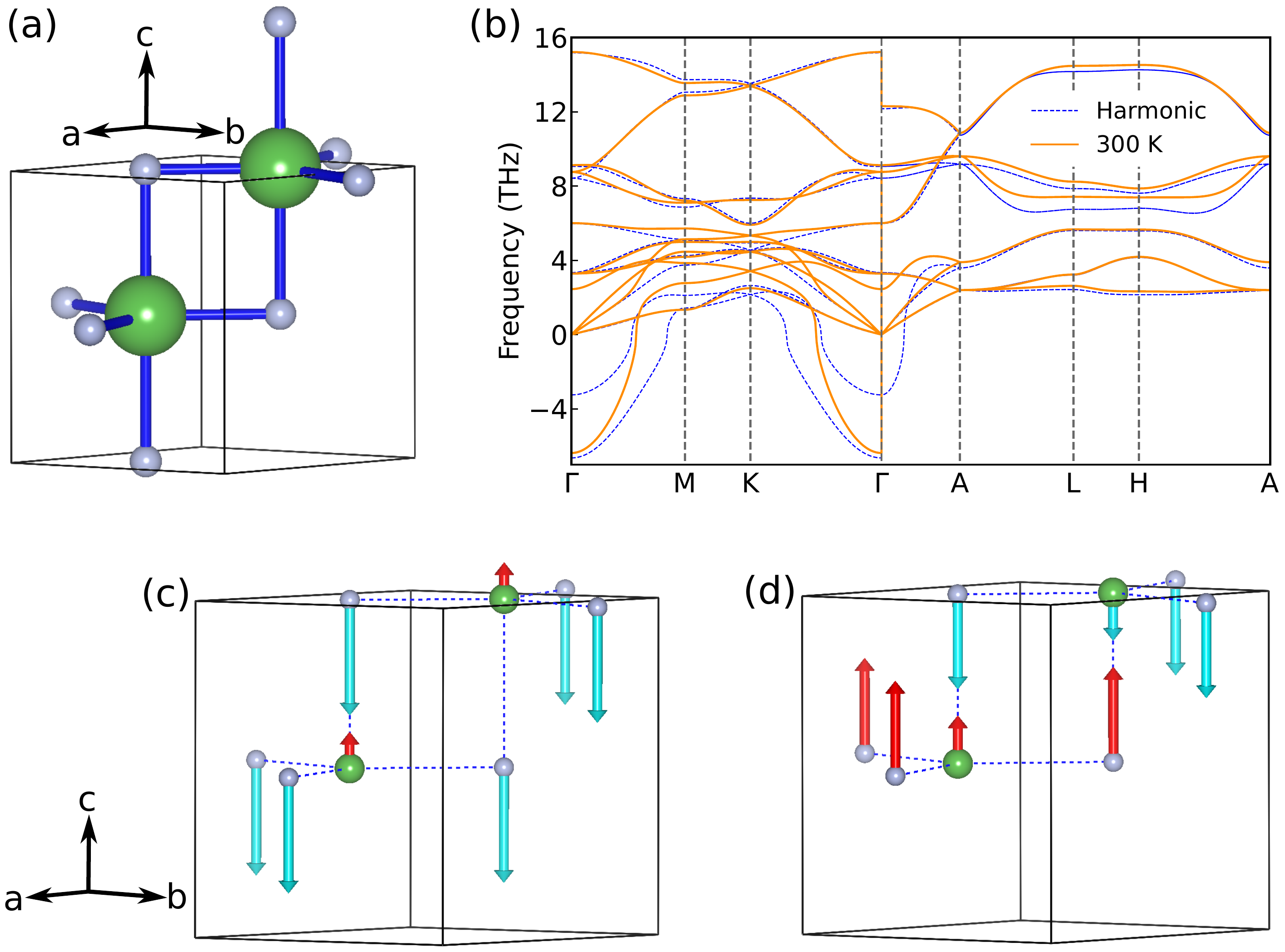}
\caption{\label{fig:Fig1} LaN-H5: (a) Unit cell containing La (green) and N (gray). (b) Phonon dispersion (harmonic approximation and at 300 K) of the H5 phase in the hexagonal Brillouin zone. Phonon eigenvectors represented by arrows on each atoms corresponding to the imaginary phonon frequency (c) -6.64 THz and (d) -3.25 at $\Gamma$.}
\end{figure}

\begin{figure}[t]
\includegraphics[width=8.3cm]{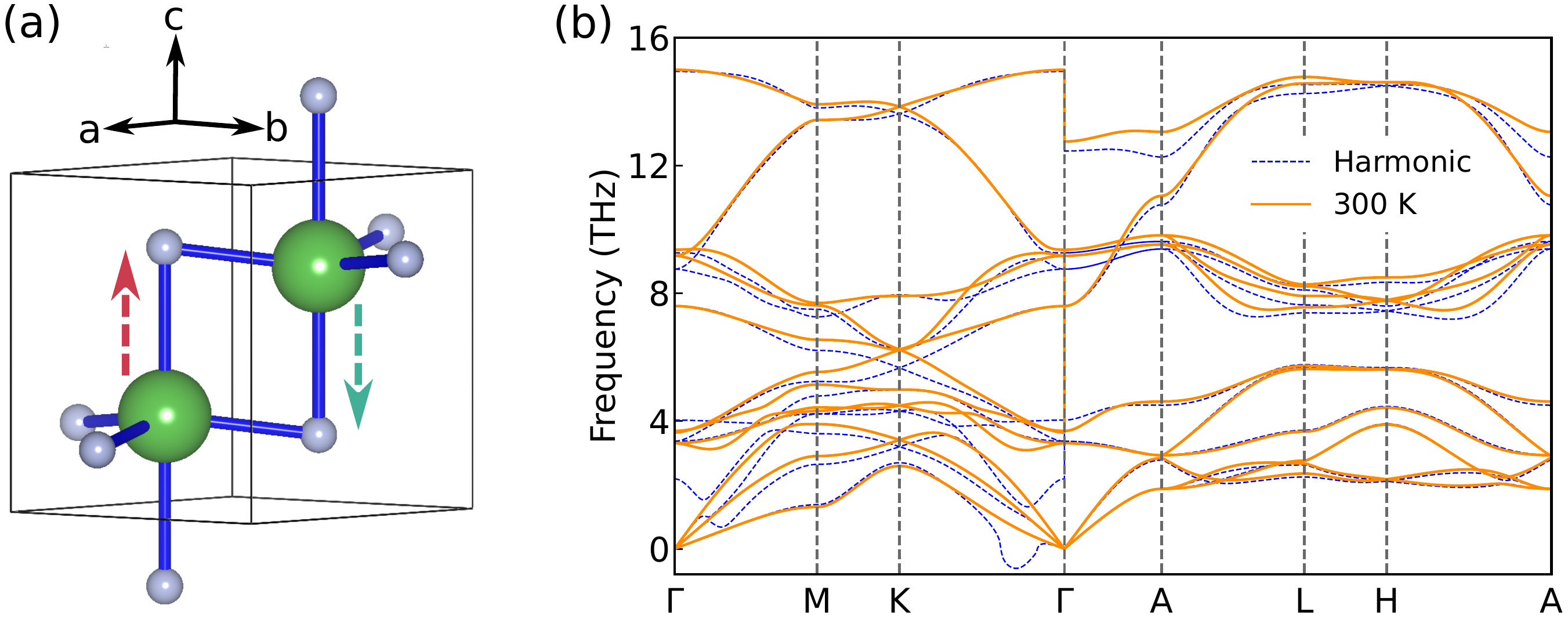}
\caption{\label{fig:Fig2}LaN-AP: (a) Unit cell containing La (green) and N (gray). Arrows indicating the direction of polarization in two sublattices. (b) Phonon dispersion of the AP phase under harmonic approximation and at 300 K in the hexagonal Brillouin zone .}
\end{figure}

In this study, we considered three hexagonal phases of LaN, each containing two formula units per cell: (1) the H5 (Fig.~\ref{fig:Fig1}) (a), (2) the WZ (inset of Fig.~\ref{fig:Fig4}(c)), and (3) the AP phase (Fig.~\ref{fig:Fig2}(a)). In the H5 phase, the La and N atoms occupy centrosymmetric positions within the $ab$ plane and overlap along the $c$-axis, preserving inversion symmetry and resulting in a nonpolar structure. In contrast, the WZ phase exhibits relative axial displacements of the La and N sublattices along the $c$-axis, breaking inversion symmetry and inducing a net macroscopic polarization.

In the AP phase, consecutive LaN layers exhibit equal and opposite relative axial displacements along the $c$-axis. In one LaN layer, the La atoms are shifted along the positive $c$ direction with respect to the neighboring N atoms, generating an upward (UP) polarization sublattice, while the adjacent LaN layer hosts a counter-shifted downward (DOWN)  sublattice. These opposing local polarizations cancel exactly, resulting in zero net macroscopic polarization within the unit cell.

\begin{figure}[b]
\includegraphics[width=8.3cm]{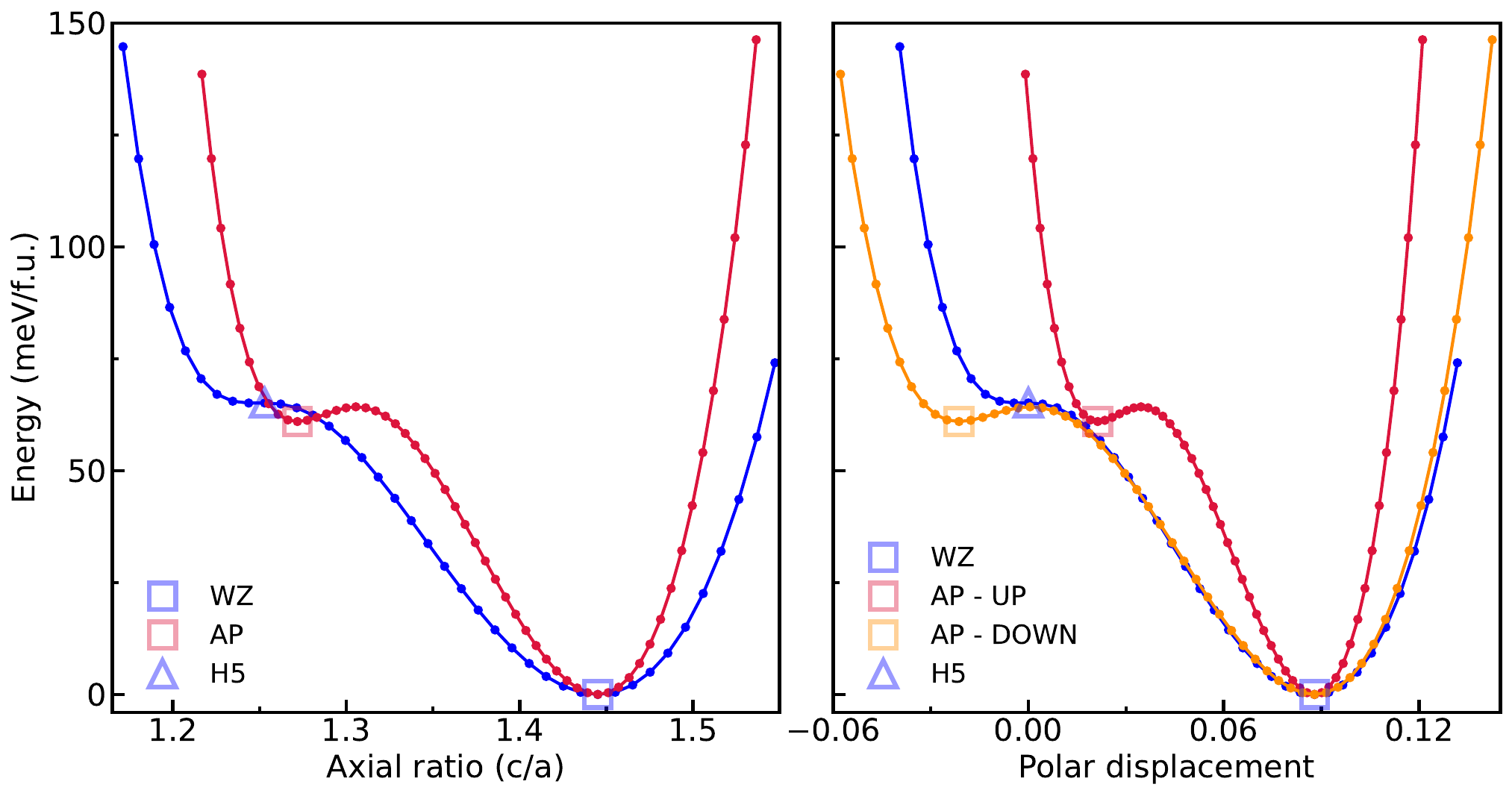}
\caption{\label{fig:Fig3} (a) Energy with axial ratio for the linearly interpolated structures between AP and WZ (red), and between H5 and WZ (blue). The H5 (blue triangle), WZ (blue square) and AP (red square) phases are indicated. (b) Energy with polar displacement for the structures in (a). Polar displacements of the WZ (blue), AP - UP sublattice (red), AP - DOWN sublattice (orange) are shown. The H5 (blue triangle), WZ (blue square), AP - UP (red square) and AP - DOWN phases are marked.}
\end{figure}

\begin{figure}[t]
\includegraphics[width=8.3cm]{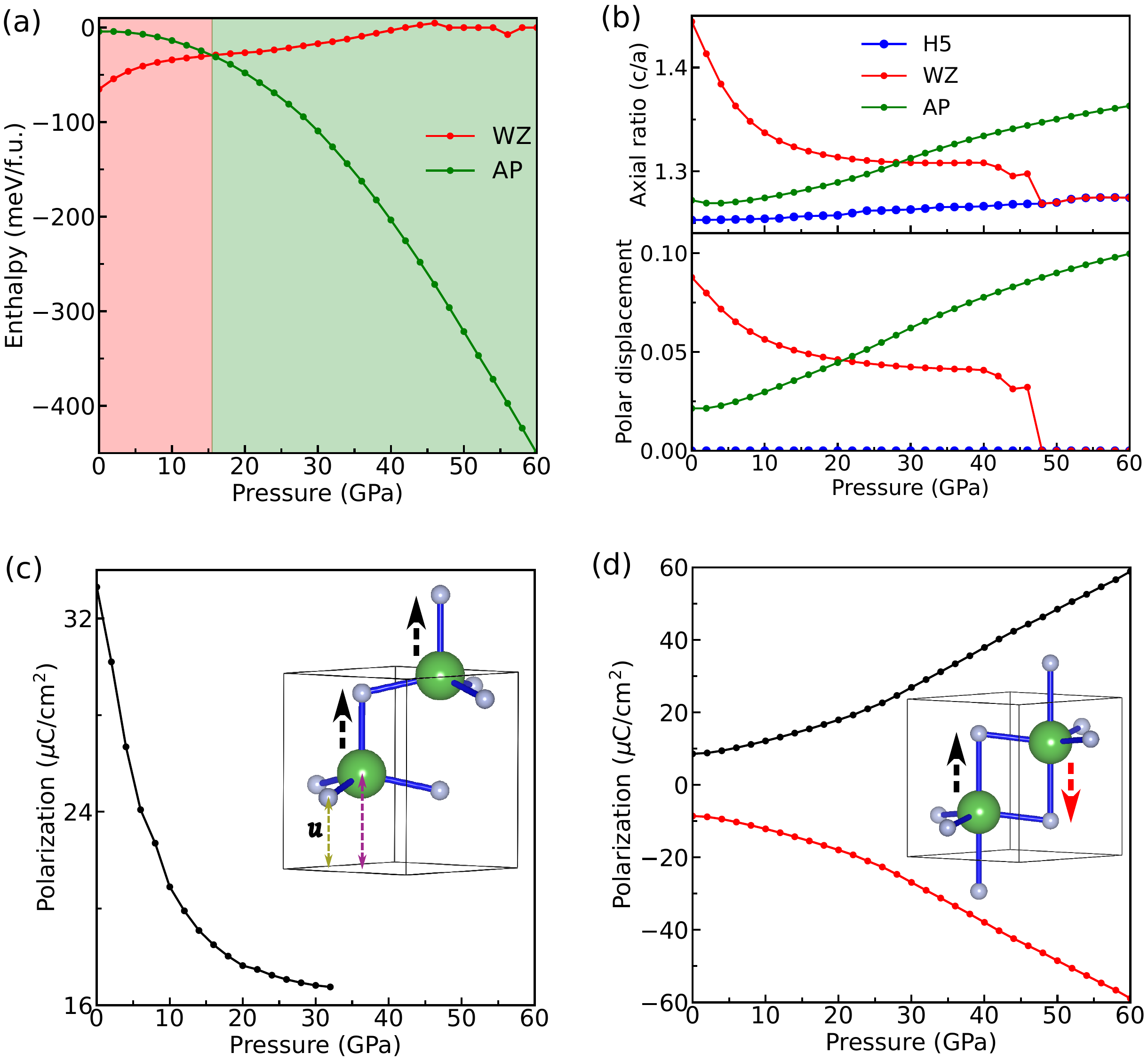}
\caption{\label{fig:Fig4} (a) Enthalpy of the WZ and AP phases with applied pressure. Zero of the enthalpy is set at the enthalpy of the H5 phase. (b) Pressure dependence of the axial ratio (top panel) and polar displacement (bottom panel). (c) Polarization contributed by each LaN\textsubscript{4} tetrahedral unit in the WZ unit cell. Inset shows the WZ structure, indicating the internal parameter $u$. (d) Polarization contributions from the up and down sublattices in AP phase. }
\end{figure}

To begin, we calculated total energies for structures obtained by linear interpolation between the H5 and WZ phases of LaN. The resulting energy path (Fig.~\ref{fig:Fig3}) exhibits a clear minimum at an axial ratio ($c/a$) of 1.45 and a polar displacement of 0.09, corresponding to the WZ phase. The polar displacement describes the relative shift of La layers with respect to neighboring N layers along the $c$ axis, which in the conventional wurtzite description corresponds to an internal parameter $u = 0.41$ (inset of Fig.~\ref{fig:Fig4}(c)). This suggests that the WZ phase can be dynamically stable and potentially experimentally realizable~\cite{Krause_nb5222}. The energy profile also shows a saddle point occurs at $c/a = 1.25$ with no polar displacement, corresponding to the H5 phase (see Supplemental Material~\cite{SM}). To assess its stability, we computed the phonon dispersion of H5 (Fig.~\ref{fig:Fig1}(b)) at both the harmonic and anharmonic levels. Within the harmonic approximation, two zone-center imaginary modes are found at -$6.64$ and -$3.25$ THz, indicating dynamical instabilities. The corresponding eigenvectors (Figs.~\ref{fig:Fig1}(c),(d)) reveal a polar and an antipolar distortion, respectively, where arrows indicate atomic displacement directions that stabilize each mode. At finite temperature, anharmonic effects stabilize the antipolar distortion, while the polar instability remains robust in the H5 structure even at 300 K. The coexistence of polar and antipolar instabilities at the zone center has also been reported in pyroxene-like oxides in the context of antiferroelectric design~\cite{Aramberri2020}.



The newly identified AP phase of LaN crystallizes in the $P\bar{3}m1$ (No. 164) space group. The structural and electronic properties of the AP phase, including comparisons with H5 and WZ phases from theory and available experiments, are provided in the Supplemental Material~\cite{SM}. The calculated axial ratio of the AP phase (1.28) lies between that of the H5 (1.25) and WZ phase (1.45). This trend reflects the nature of the
underlying distortions: the WZ structure promotes a net polar distortion along the $c$-axis, leading to a larger axial ratio. In contrast, the AP phase consists of locally polar LaN layers arranged in an alternating manner, yielding zero net polarization. Consequently, the axial ratio of the AP phase remains closer to that of the nonpolar H5 phase.

It is noteworthy that the axial ratio of the WZ phase of LaN (1.45) is significantly smaller than the typical value ($\sim$ 1.6) observed in other group III-V nitrides such as AlN, GaN, and InN. This suggests a comparatively weaker polar distortion in WZ LaN and may explain the absence of an unstable AP mode in these other polar nitrides (calculations not shown here).

From total energy calculations, the WZ phase is found to be the most stable among the considered hexagonal phases of LaN. The AP phase is 62 meV/f.u. higher than the WZ phase, indicating that the AP phase is energetically competitive and potentially metastable. Meanwhile the H5 phase lies 68 meV/f.u. above the WZ phase.

To quantify the polar distortion, we calculate the relative axial displacement of La layers with respect to neighboring N layers, yielding 0.53 \AA~ for the WZ phase and 0.12 \AA~ for the AP phase. The spontaneous polarization can then be estimated from atomic displacements relative to their centrosymmetric positions and the corresponding Born effective charges (see Supplemental Material~\cite{SM}). Assuming a linear relationship between polarization change and atomic displacement, the contribution from each atomic displacement to the total polarization of the unit cell can be expressed as~\cite{Born_P1,Resta},
\begin{equation}
\Delta P_{i} = \frac{e}{\Omega} \displaystyle \sum_{\kappa j} Z^{*}_{\kappa ij}\Delta d_{\kappa j}
\end{equation}
where $e$ is the electron charge and $\Omega$ is the unit cell volume. $Z^{*}_{\kappa ij}$ denotes the Born effective charge tensor of the $\kappa$-th atom, and $\Delta \boldsymbol{d}_{\kappa}$ represents the displacement vector of the atom from its centrosymmetric reference position. The indices $i$ and $j$ correspond to Cartesian directions. 

Using this formalism, each sublattice of the AP phase contributes $\approx$ 9.00 $\mu$C/cm$^{2}$, which cancels globally due to the antipolar layer stacking. In contrast, each sublattice in the WZ phase contributes $\approx$ 33.50 $\mu$C/cm$^{2}$, yielding a total spontaneous macroscopic polarization of $\approx$ 67.00 $\mu$C/cm$^{2}$ along the \(c\) axis.

Following the structural prediction of the AP phase, we examined its dynamical stability by calculating the phonon dispersion in the hexagonal unit cell considering harmonic and anharmonic limit, as shown in Fig.~\ref{fig:Fig2}(b). Though, in the harmonic limit the phonon dispersion shows a small imaginary frequency along the K-$\Gamma$. It becomes positive while considering the anharmonic effect at 300 K. The absence of imaginary frequencies throughout the Brillouin zone confirms the dynamical stability of the AP phase.




To further assess the stability of the AP phase, we examined the calculated energy path connecting the AP and WZ phases, as shown in Fig.~\ref{fig:Fig3}. Figure~\ref{fig:Fig3}(a) demonstrates that, unlike the energy path between the H5 and WZ phases, the AP-WZ energy path exhibits two distinct minima. One minimum occurs at an axial ratio of 1.45, corresponding to the WZ phase, while the other appears at 1.28, corresponding to the AP phase. As shown in Fig.~\ref{fig:Fig3}(b), the presence of energy minima at polar displacements of 0.02 and -0.02 corresponds to stable polar displacements in the UP and DOWN sublattices along the transformation path between the AP and WZ structures.


Although the total energy of the H5 phase is close to that of the AP phase under ambient conditions (AP is 6 meV/f.u. lower in energy compared to H5), the presence of imaginary phonon modes confirms its dynamical instability. Therefore, if WZ LaN becomes ferroelectric, polarization switching under an external electric field is unlikely to proceed through the H5 structure, despite its comparable energy, in contrast to a recent theoretical study~\cite{PhysRevMaterials.5.094602}. Instead, the switching pathway is more likely to follow dynamically stable distortions, potentially involving the antipolar (AP) phase.

To gain further insight into the switching pathway, we decomposed the total energy into ionic, elastic, and coupling contributions (see Supplemental Material~\cite{SM}). This analysis reveals that the switching energetics are primarily driven by ionic displacements, with lattice deformation substantially renormalizing the energy landscape.

It should be noted that field-induced polarization switching in real materials involves complex extrinsic processes-such as domain nucleation and domain wall motion-alongside intrinsic effects~\cite{Zheng2026, Yazawa2023}, a detailed investigation of these kinetics is beyond the scope of this work. Here, we restrict our discussion to the intrinsic energy landscape obtained from first-principles calculations.

Importantly, the calculated energy barrier between the WZ and AP phases is significantly smaller than the reported switching barriers between the WZ and H5 phases in bulk AlN ($\sim$ 230 meV/f.u.) and Al$_{1-x}$Sc$_{x}$N ($x = 0.3$) ($\sim$ 100 meV/f.u.)~\cite{AlN_height}. This comparatively lower barrier suggests that polarization switching in LaN may require a smaller electric field, potentially well below the dielectric breakdown limit of the material.

The significant differences of c/a ratio between the AP and WZ phases indicates (1) the presence of strong electromechanical couplings, as recently revealed in (Al,Sc)N~\cite{Peng} 
and (2) that mechanical constraints such as hydrostatic pressure or uniaxial strain may facilitate the emergence of the AP phase. Here, we focus on the effect of hydrostatic pressure on the H5, WZ, and AP phases in order to assess the experimental feasibility of realizing the AP phase. Figure~\ref{fig:Fig4}(a) shows the calculated enthalpy ($H = E + PV$) of these phases as a function of pressure. Although the AP phase is higher in energy than the WZ phase at ambient conditions, its enthalpy decreases relative to WZ with increasing pressure, while the enthalpy of the WZ phase increases, indicating a pressure-induced stabilization of the AP phase with respect to the WZ phase. The AP phase becomes energetically more favorable than the WZ phase above $\approx$ 15.5 GPa. The pressure-induced stabilization of the AP phase originates primarily from its more efficient volume reduction under compression. While the internal energy contribution does not favor AP over WZ, the enhanced antipolar distortions enable greater structural accommodation of compression, leading to a smaller volume, lower $PV$ contribution, and consequently lower enthalpy at high pressure (see Supplemental Material~\cite{SM}). However, when considering other competing phases of LaN under hydrostatic pressure, a different picture emerges (see Supplemental Material~\cite{SM}). The enthalpies of the RS, $P1$, and tetragonal phases decrease more rapidly with pressure. In particular, the tetragonal phase becomes the lowest-enthalpy structure above 20 GPa, consistent with experimental observations that this phase is synthesized under high-pressure conditions~\cite{LaN_press}. Consequently, although the AP phase becomes more stable than WZ at intermediate pressures, it does not represent the global ground state under high pressure. Notably, the enthalpy difference between the tetragonal and AP phases increases rapidly with pressure, further reducing the possibility of stabilizing the AP phase under high-pressure conditions.

These results suggest that realizing the AP phase under very high pressure is unlikely. However, in the low-to-intermediate pressure range, the enthalpy differences among the competing phases remain relatively small and tunable. We therefore propose that the AP phase may be experimentally accessible in this moderate pressure window. If stabilized, the energy barrier between the AP and WZ phases could be tuned by pressure, potentially enabling the realization of antiferroelectric switching in LaN.

Figure~\ref{fig:Fig4}(b) shows the evolution of the axial ratio and polar displacement under pressure for the H5, WZ, and AP phases. In the WZ phase, both quantities decrease with increasing pressure, indicating suppression of the polar distortion. Above 40 GPa, the WZ phase gradually approaches the H5 structure, consistent with the reduction of its polar character. In contrast, for the AP phase, both the axial ratio and the magnitude of the local polar displacement increase with pressure. This suggests that the applied pressure is accommodated by enhancing the local polar distortions in opposite directions within the two La-N sublattices of the unit cell, thereby preserving the overall antipolar arrangement.

Figures~\ref{fig:Fig4}(c) and (d) present the calculated polarization of each LaN$_4$ tetrahedral sublattice in the WZ and AP phases, respectively. The total polarization of the WZ phase is twice the value shown in Fig.~\ref{fig:Fig4}(c), since both sublattices contribute constructively. As expected, it decreases with increasing pressure, consistent with the reduction in polar displacement of the WZ phase. In contrast, the magnitude of the local polarization in each sublattice of the AP phase increases with pressure, while the net macroscopic polarization remains zero due to antipolar stacking.



In summary, the nonpolar H5 phase of hexagonal LaN is dynamically unstable, exhibiting both polar and antipolar zone-center lattice instabilities. Condensation of these instabilities stabilizes, respectively, a WZ phase and an AP phase. To the best of our knowledge, this represents the first prediction of an antipolar instability in a hexagonal binary nitride. The AP phase is characterized by alternating local polarization with zero net macroscopic polarization and is metastable with respect to the WZ phase at ambient conditions. A finite energy barrier between the WZ and AP phases suggests a possible polarization-switching pathway under an applied electric field, mediated by the AP phase.


We show that hydrostatic pressure significantly tunes the competition between these phases. In particular, the energy difference between the WZ and AP phases decreases with increasing pressure in the low-pressure regime, while the sublattice polarization increases in the AP phase and decreases in the WZ phase. These contrasting trends reflect the fundamentally different nature of polar and antipolar distortions in LaN. At higher pressures, however, competing rocksalt and tetragonal phases become energetically more favorable, indicating that the AP phase is most likely to be realized in the low-pressure regime. Overall, our results reveal a pressure-tunable interplay between polar and antipolar states in a binary nitride and suggest the possible stabilization of ferroelectric and antiferroelectric-like behaviors in LaN. This AP phase in LaN may be one of the few examples of a potential proper antiferroelectric phase, in contrast to perovskite oxides in which most antiferroelectric phases are mediated by or hybridized with oxygen tilt octahedra~\cite{Laurent,Kinnary}.

\textit{Acknowledgments---}This work is supported by a grant from the U.S. Department of Energy under award no. DE-SC0025479.  We also thank Dr. Ashis Kundu for useful discussions on finite-temperature phonon calculations. This work used computational resources provided by the Arkansas High Performance Computing Center, which is funded through multiple National Science Foundation grants and the Arkansas Economic Development Commission.

\bibliography{LaN_AP}

\providecommand{\noopsort}[1]{}\providecommand{\singleletter}[1]{#1}%
\begin{thebibliography}{48}%
\makeatletter
\providecommand \@ifxundefined [1]{%
 \@ifx{#1\undefined}
}%
\providecommand \@ifnum [1]{%
 \ifnum #1\expandafter \@firstoftwo
 \else \expandafter \@secondoftwo
 \fi
}%
\providecommand \@ifx [1]{%
 \ifx #1\expandafter \@firstoftwo
 \else \expandafter \@secondoftwo
 \fi
}%
\providecommand \natexlab [1]{#1}%
\providecommand \enquote  [1]{``#1''}%
\providecommand \bibnamefont  [1]{#1}%
\providecommand \bibfnamefont [1]{#1}%
\providecommand \citenamefont [1]{#1}%
\providecommand \href@noop [0]{\@secondoftwo}%
\providecommand \href [0]{\begingroup \@sanitize@url \@href}%
\providecommand \@href[1]{\@@startlink{#1}\@@href}%
\providecommand \@@href[1]{\endgroup#1\@@endlink}%
\providecommand \@sanitize@url [0]{\catcode `\\12\catcode `\$12\catcode
  `\&12\catcode `\#12\catcode `\^12\catcode `\_12\catcode `\%12\relax}%
\providecommand \@@startlink[1]{}%
\providecommand \@@endlink[0]{}%
\providecommand \url  [0]{\begingroup\@sanitize@url \@url }%
\providecommand \@url [1]{\endgroup\@href {#1}{\urlprefix }}%
\providecommand \urlprefix  [0]{URL }%
\providecommand \Eprint [0]{\href }%
\providecommand \doibase [0]{https://doi.org/}%
\providecommand \selectlanguage [0]{\@gobble}%
\providecommand \bibinfo  [0]{\@secondoftwo}%
\providecommand \bibfield  [0]{\@secondoftwo}%
\providecommand \translation [1]{[#1]}%
\providecommand \BibitemOpen [0]{}%
\providecommand \bibitemStop [0]{}%
\providecommand \bibitemNoStop [0]{.\EOS\space}%
\providecommand \EOS [0]{\spacefactor3000\relax}%
\providecommand \BibitemShut  [1]{\csname bibitem#1\endcsname}%
\let\auto@bib@innerbib\@empty
\bibitem [{\citenamefont {Nakamura}(2015)}]{RevModPhys.87.1139}%
  \BibitemOpen
  \bibfield  {author} {\bibinfo {author} {\bibfnamefont {S.}~\bibnamefont
  {Nakamura}},\ }\bibfield  {title} {\bibinfo {title} {Nobel lecture:
  Background story of the invention of efficient blue ingan light emitting
  diodes},\ }\href {https://doi.org/10.1103/RevModPhys.87.1139} {\bibfield
  {journal} {\bibinfo  {journal} {Rev. Mod. Phys.}\ }\textbf {\bibinfo {volume}
  {87}},\ \bibinfo {pages} {1139} (\bibinfo {year} {2015})}\BibitemShut
  {NoStop}%
\bibitem [{\citenamefont {Akasaki}(2015)}]{RevModPhys.87.1119}%
  \BibitemOpen
  \bibfield  {author} {\bibinfo {author} {\bibfnamefont {I.}~\bibnamefont
  {Akasaki}},\ }\bibfield  {title} {\bibinfo {title} {Nobel lecture: Fascinated
  journeys into blue light},\ }\href
  {https://doi.org/10.1103/RevModPhys.87.1119} {\bibfield  {journal} {\bibinfo
  {journal} {Rev. Mod. Phys.}\ }\textbf {\bibinfo {volume} {87}},\ \bibinfo
  {pages} {1119} (\bibinfo {year} {2015})}\BibitemShut {NoStop}%
\bibitem [{\citenamefont {Liu}\ \emph {et~al.}(2026)\citenamefont {Liu},
  \citenamefont {Zhu}, \citenamefont {Niroula}, \citenamefont {Pal},
  \citenamefont {Palacios},\ and\ \citenamefont {Eisner}}]{Liu2026}%
  \BibitemOpen
  \bibfield  {author} {\bibinfo {author} {\bibfnamefont {Y.-C.}\ \bibnamefont
  {Liu}}, \bibinfo {author} {\bibfnamefont {J.}~\bibnamefont {Zhu}}, \bibinfo
  {author} {\bibfnamefont {J.}~\bibnamefont {Niroula}}, \bibinfo {author}
  {\bibfnamefont {H.}~\bibnamefont {Pal}}, \bibinfo {author} {\bibfnamefont
  {T.}~\bibnamefont {Palacios}},\ and\ \bibinfo {author} {\bibfnamefont
  {S.~R.}\ \bibnamefont {Eisner}},\ }\bibfield  {title} {\bibinfo {title}
  {High-temperature operation of group-iii nitride high-electron-mobility
  transistors},\ }\bibfield  {journal} {\bibinfo  {journal} {Nature
  Electronics}\ }\href {https://doi.org/10.1038/s41928-026-01570-y}
  {10.1038/s41928-026-01570-y} (\bibinfo {year} {2026})\BibitemShut {NoStop}%
\bibitem [{\citenamefont {Hoque}\ \emph {et~al.}(2021)\citenamefont {Hoque},
  \citenamefont {Koh}, \citenamefont {Braun}, \citenamefont {Mamun},
  \citenamefont {Liu}, \citenamefont {Huynh}, \citenamefont {Liao},
  \citenamefont {Hussain}, \citenamefont {Cheng}, \citenamefont {Hoglund},
  \citenamefont {Olson}, \citenamefont {Tomko}, \citenamefont {Aryana},
  \citenamefont {Galib}, \citenamefont {Gaskins}, \citenamefont {Elahi},
  \citenamefont {Leseman}, \citenamefont {Howe}, \citenamefont {Luo},
  \citenamefont {Graham}, \citenamefont {Goorsky}, \citenamefont {Khan},\ and\
  \citenamefont {Hopkins}}]{Thermal_Conduct}%
  \BibitemOpen
  \bibfield  {author} {\bibinfo {author} {\bibfnamefont {M.~S.~B.}\
  \bibnamefont {Hoque}}, \bibinfo {author} {\bibfnamefont {Y.~R.}\ \bibnamefont
  {Koh}}, \bibinfo {author} {\bibfnamefont {J.~L.}\ \bibnamefont {Braun}},
  \bibinfo {author} {\bibfnamefont {A.}~\bibnamefont {Mamun}}, \bibinfo
  {author} {\bibfnamefont {Z.}~\bibnamefont {Liu}}, \bibinfo {author}
  {\bibfnamefont {K.}~\bibnamefont {Huynh}}, \bibinfo {author} {\bibfnamefont
  {M.~E.}\ \bibnamefont {Liao}}, \bibinfo {author} {\bibfnamefont
  {K.}~\bibnamefont {Hussain}}, \bibinfo {author} {\bibfnamefont
  {Z.}~\bibnamefont {Cheng}}, \bibinfo {author} {\bibfnamefont {E.~R.}\
  \bibnamefont {Hoglund}}, \bibinfo {author} {\bibfnamefont {D.~H.}\
  \bibnamefont {Olson}}, \bibinfo {author} {\bibfnamefont {J.~A.}\ \bibnamefont
  {Tomko}}, \bibinfo {author} {\bibfnamefont {K.}~\bibnamefont {Aryana}},
  \bibinfo {author} {\bibfnamefont {R.}~\bibnamefont {Galib}}, \bibinfo
  {author} {\bibfnamefont {J.~T.}\ \bibnamefont {Gaskins}}, \bibinfo {author}
  {\bibfnamefont {M.~M.~M.}\ \bibnamefont {Elahi}}, \bibinfo {author}
  {\bibfnamefont {Z.~C.}\ \bibnamefont {Leseman}}, \bibinfo {author}
  {\bibfnamefont {J.~M.}\ \bibnamefont {Howe}}, \bibinfo {author}
  {\bibfnamefont {T.}~\bibnamefont {Luo}}, \bibinfo {author} {\bibfnamefont
  {S.}~\bibnamefont {Graham}}, \bibinfo {author} {\bibfnamefont {M.~S.}\
  \bibnamefont {Goorsky}}, \bibinfo {author} {\bibfnamefont {A.}~\bibnamefont
  {Khan}},\ and\ \bibinfo {author} {\bibfnamefont {P.~E.}\ \bibnamefont
  {Hopkins}},\ }\bibfield  {title} {\bibinfo {title} {High in-plane thermal
  conductivity of aluminum nitride thin films},\ }\href
  {https://doi.org/10.1021/acsnano.0c09915} {\bibfield  {journal} {\bibinfo
  {journal} {ACS Nano}\ }\textbf {\bibinfo {volume} {15}},\ \bibinfo {pages}
  {9588} (\bibinfo {year} {2021})},\ \bibinfo {note} {pMID:
  33908771}\BibitemShut {NoStop}%
\bibitem [{\citenamefont {Amano}\ \emph {et~al.}(2020)\citenamefont {Amano},
  \citenamefont {Collazo}, \citenamefont {Santi}, \citenamefont {Einfeldt},
  \citenamefont {Funato}, \citenamefont {Glaab}, \citenamefont {Hagedorn},
  \citenamefont {Hirano}, \citenamefont {Hirayama}, \citenamefont {Ishii},
  \citenamefont {Kashima}, \citenamefont {Kawakami}, \citenamefont {Kirste},
  \citenamefont {Kneissl}, \citenamefont {Martin}, \citenamefont {Mehnke},
  \citenamefont {Meneghini}, \citenamefont {Ougazzaden}, \citenamefont
  {Parbrook}, \citenamefont {Rajan}, \citenamefont {Reddy}, \citenamefont
  {Römer}, \citenamefont {Ruschel}, \citenamefont {Sarkar}, \citenamefont
  {Scholz}, \citenamefont {Schowalter}, \citenamefont {Shields}, \citenamefont
  {Sitar}, \citenamefont {Sulmoni}, \citenamefont {Wang}, \citenamefont
  {Wernicke}, \citenamefont {Weyers}, \citenamefont {Witzigmann}, \citenamefont
  {Wu}, \citenamefont {Wunderer},\ and\ \citenamefont {Zhang}}]{Amano_2020}%
  \BibitemOpen
  \bibfield  {author} {\bibinfo {author} {\bibfnamefont {H.}~\bibnamefont
  {Amano}}, \bibinfo {author} {\bibfnamefont {R.}~\bibnamefont {Collazo}},
  \bibinfo {author} {\bibfnamefont {C.~D.}\ \bibnamefont {Santi}}, \bibinfo
  {author} {\bibfnamefont {S.}~\bibnamefont {Einfeldt}}, \bibinfo {author}
  {\bibfnamefont {M.}~\bibnamefont {Funato}}, \bibinfo {author} {\bibfnamefont
  {J.}~\bibnamefont {Glaab}}, \bibinfo {author} {\bibfnamefont
  {S.}~\bibnamefont {Hagedorn}}, \bibinfo {author} {\bibfnamefont
  {A.}~\bibnamefont {Hirano}}, \bibinfo {author} {\bibfnamefont
  {H.}~\bibnamefont {Hirayama}}, \bibinfo {author} {\bibfnamefont
  {R.}~\bibnamefont {Ishii}}, \bibinfo {author} {\bibfnamefont
  {Y.}~\bibnamefont {Kashima}}, \bibinfo {author} {\bibfnamefont
  {Y.}~\bibnamefont {Kawakami}}, \bibinfo {author} {\bibfnamefont
  {R.}~\bibnamefont {Kirste}}, \bibinfo {author} {\bibfnamefont
  {M.}~\bibnamefont {Kneissl}}, \bibinfo {author} {\bibfnamefont
  {R.}~\bibnamefont {Martin}}, \bibinfo {author} {\bibfnamefont
  {F.}~\bibnamefont {Mehnke}}, \bibinfo {author} {\bibfnamefont
  {M.}~\bibnamefont {Meneghini}}, \bibinfo {author} {\bibfnamefont
  {A.}~\bibnamefont {Ougazzaden}}, \bibinfo {author} {\bibfnamefont {P.~J.}\
  \bibnamefont {Parbrook}}, \bibinfo {author} {\bibfnamefont {S.}~\bibnamefont
  {Rajan}}, \bibinfo {author} {\bibfnamefont {P.}~\bibnamefont {Reddy}},
  \bibinfo {author} {\bibfnamefont {F.}~\bibnamefont {Römer}}, \bibinfo
  {author} {\bibfnamefont {J.}~\bibnamefont {Ruschel}}, \bibinfo {author}
  {\bibfnamefont {B.}~\bibnamefont {Sarkar}}, \bibinfo {author} {\bibfnamefont
  {F.}~\bibnamefont {Scholz}}, \bibinfo {author} {\bibfnamefont {L.~J.}\
  \bibnamefont {Schowalter}}, \bibinfo {author} {\bibfnamefont
  {P.}~\bibnamefont {Shields}}, \bibinfo {author} {\bibfnamefont
  {Z.}~\bibnamefont {Sitar}}, \bibinfo {author} {\bibfnamefont
  {L.}~\bibnamefont {Sulmoni}}, \bibinfo {author} {\bibfnamefont
  {T.}~\bibnamefont {Wang}}, \bibinfo {author} {\bibfnamefont {T.}~\bibnamefont
  {Wernicke}}, \bibinfo {author} {\bibfnamefont {M.}~\bibnamefont {Weyers}},
  \bibinfo {author} {\bibfnamefont {B.}~\bibnamefont {Witzigmann}}, \bibinfo
  {author} {\bibfnamefont {Y.-R.}\ \bibnamefont {Wu}}, \bibinfo {author}
  {\bibfnamefont {T.}~\bibnamefont {Wunderer}},\ and\ \bibinfo {author}
  {\bibfnamefont {Y.}~\bibnamefont {Zhang}},\ }\bibfield  {title} {\bibinfo
  {title} {The 2020 uv emitter roadmap},\ }\href
  {https://doi.org/10.1088/1361-6463/aba64c} {\bibfield  {journal} {\bibinfo
  {journal} {Journal of Physics D: Applied Physics}\ }\textbf {\bibinfo
  {volume} {53}},\ \bibinfo {pages} {503001} (\bibinfo {year}
  {2020})}\BibitemShut {NoStop}%
\bibitem [{\citenamefont {Kudrawiec}\ and\ \citenamefont
  {Hommel}(2020)}]{band_Gap_Engi}%
  \BibitemOpen
  \bibfield  {author} {\bibinfo {author} {\bibfnamefont {R.}~\bibnamefont
  {Kudrawiec}}\ and\ \bibinfo {author} {\bibfnamefont {D.}~\bibnamefont
  {Hommel}},\ }\bibfield  {title} {\bibinfo {title} {Bandgap engineering in
  iii-nitrides with boron and group v elements: Toward applications in
  ultraviolet emitters},\ }\href {https://doi.org/10.1063/5.0025371} {\bibfield
   {journal} {\bibinfo  {journal} {Applied Physics Reviews}\ }\textbf {\bibinfo
  {volume} {7}},\ \bibinfo {pages} {041314} (\bibinfo {year}
  {2020})}\BibitemShut {NoStop}%
\bibitem [{\citenamefont {Zhao}\ \emph {et~al.}(2015)\citenamefont {Zhao},
  \citenamefont {Nguyen}, \citenamefont {Kibria},\ and\ \citenamefont
  {Mi}}]{ZHAO201514}%
  \BibitemOpen
  \bibfield  {author} {\bibinfo {author} {\bibfnamefont {S.}~\bibnamefont
  {Zhao}}, \bibinfo {author} {\bibfnamefont {H.~P.}\ \bibnamefont {Nguyen}},
  \bibinfo {author} {\bibfnamefont {M.~G.}\ \bibnamefont {Kibria}},\ and\
  \bibinfo {author} {\bibfnamefont {Z.}~\bibnamefont {Mi}},\ }\bibfield
  {title} {\bibinfo {title} {Iii-nitride nanowire optoelectronics},\ }\href
  {https://doi.org/https://doi.org/10.1016/j.pquantelec.2015.11.001} {\bibfield
   {journal} {\bibinfo  {journal} {Progress in Quantum Electronics}\ }\textbf
  {\bibinfo {volume} {44}},\ \bibinfo {pages} {14} (\bibinfo {year}
  {2015})}\BibitemShut {NoStop}%
\bibitem [{\citenamefont {Fichtner}\ \emph {et~al.}(2019)\citenamefont
  {Fichtner}, \citenamefont {Wolff}, \citenamefont {Lofink}, \citenamefont
  {Kienle},\ and\ \citenamefont {Wagner}}]{AlSCN_switch}%
  \BibitemOpen
  \bibfield  {author} {\bibinfo {author} {\bibfnamefont {S.}~\bibnamefont
  {Fichtner}}, \bibinfo {author} {\bibfnamefont {N.}~\bibnamefont {Wolff}},
  \bibinfo {author} {\bibfnamefont {F.}~\bibnamefont {Lofink}}, \bibinfo
  {author} {\bibfnamefont {L.}~\bibnamefont {Kienle}},\ and\ \bibinfo {author}
  {\bibfnamefont {B.}~\bibnamefont {Wagner}},\ }\bibfield  {title} {\bibinfo
  {title} {Alscn: A iii-v semiconductor based ferroelectric},\ }\href
  {https://doi.org/10.1063/1.5084945} {\bibfield  {journal} {\bibinfo
  {journal} {Journal of Applied Physics}\ }\textbf {\bibinfo {volume} {125}},\
  \bibinfo {pages} {114103} (\bibinfo {year} {2019})}\BibitemShut {NoStop}%
\bibitem [{\citenamefont {Mikolajick}\ \emph {et~al.}(2021)\citenamefont
  {Mikolajick}, \citenamefont {Slesazeck}, \citenamefont {Mulaosmanovic},
  \citenamefont {Park}, \citenamefont {Fichtner}, \citenamefont {Lomenzo},
  \citenamefont {Hoffmann},\ and\ \citenamefont {Schroeder}}]{FE_Next}%
  \BibitemOpen
  \bibfield  {author} {\bibinfo {author} {\bibfnamefont {T.}~\bibnamefont
  {Mikolajick}}, \bibinfo {author} {\bibfnamefont {S.}~\bibnamefont
  {Slesazeck}}, \bibinfo {author} {\bibfnamefont {H.}~\bibnamefont
  {Mulaosmanovic}}, \bibinfo {author} {\bibfnamefont {M.~H.}\ \bibnamefont
  {Park}}, \bibinfo {author} {\bibfnamefont {S.}~\bibnamefont {Fichtner}},
  \bibinfo {author} {\bibfnamefont {P.~D.}\ \bibnamefont {Lomenzo}}, \bibinfo
  {author} {\bibfnamefont {M.}~\bibnamefont {Hoffmann}},\ and\ \bibinfo
  {author} {\bibfnamefont {U.}~\bibnamefont {Schroeder}},\ }\bibfield  {title}
  {\bibinfo {title} {Next generation ferroelectric materials for semiconductor
  process integration and their applications},\ }\href
  {https://doi.org/10.1063/5.0037617} {\bibfield  {journal} {\bibinfo
  {journal} {Journal of Applied Physics}\ }\textbf {\bibinfo {volume} {129}},\
  \bibinfo {pages} {100901} (\bibinfo {year} {2021})}\BibitemShut {NoStop}%
\bibitem [{\citenamefont {Wang}\ \emph {et~al.}(2025)\citenamefont {Wang},
  \citenamefont {Ye}, \citenamefont {Xu}, \citenamefont {Wang}, \citenamefont
  {Feng}, \citenamefont {Wang}, \citenamefont {Sheng}, \citenamefont {Liu},
  \citenamefont {Shen}, \citenamefont {Wang},\ and\ \citenamefont
  {Wang}}]{FE_Next1}%
  \BibitemOpen
  \bibfield  {author} {\bibinfo {author} {\bibfnamefont {R.}~\bibnamefont
  {Wang}}, \bibinfo {author} {\bibfnamefont {H.}~\bibnamefont {Ye}}, \bibinfo
  {author} {\bibfnamefont {X.}~\bibnamefont {Xu}}, \bibinfo {author}
  {\bibfnamefont {J.}~\bibnamefont {Wang}}, \bibinfo {author} {\bibfnamefont
  {R.}~\bibnamefont {Feng}}, \bibinfo {author} {\bibfnamefont {T.}~\bibnamefont
  {Wang}}, \bibinfo {author} {\bibfnamefont {B.}~\bibnamefont {Sheng}},
  \bibinfo {author} {\bibfnamefont {F.}~\bibnamefont {Liu}}, \bibinfo {author}
  {\bibfnamefont {B.}~\bibnamefont {Shen}}, \bibinfo {author} {\bibfnamefont
  {P.}~\bibnamefont {Wang}},\ and\ \bibinfo {author} {\bibfnamefont
  {X.}~\bibnamefont {Wang}},\ }\bibfield  {title} {\bibinfo {title}
  {Composition-graded nitride ferroelectrics based multi-level non-volatile
  memory for neuromorphic computing},\ }\href
  {https://doi.org/https://doi.org/10.1002/adma.202414805} {\bibfield
  {journal} {\bibinfo  {journal} {Advanced Materials}\ }\textbf {\bibinfo
  {volume} {37}},\ \bibinfo {pages} {2414805} (\bibinfo {year}
  {2025})}\BibitemShut {NoStop}%
\bibitem [{\citenamefont {Catalan}\ \emph {et~al.}(2026)\citenamefont
  {Catalan}, \citenamefont {Gruverman}, \citenamefont
  {{\'I}{\~{n}}iguez-Gonz{\'a}lez}, \citenamefont {Meier},\ and\ \citenamefont
  {Trassin}}]{Catalan2026}%
  \BibitemOpen
  \bibfield  {author} {\bibinfo {author} {\bibfnamefont {G.}~\bibnamefont
  {Catalan}}, \bibinfo {author} {\bibfnamefont {A.}~\bibnamefont {Gruverman}},
  \bibinfo {author} {\bibfnamefont {J.}~\bibnamefont
  {{\'I}{\~{n}}iguez-Gonz{\'a}lez}}, \bibinfo {author} {\bibfnamefont
  {D.}~\bibnamefont {Meier}},\ and\ \bibinfo {author} {\bibfnamefont
  {M.}~\bibnamefont {Trassin}},\ }\bibfield  {title} {\bibinfo {title} {A
  modern perspective on antiferroelectrics},\ }\href
  {https://doi.org/10.1038/s41563-026-02483-z} {\bibfield  {journal} {\bibinfo
  {journal} {Nature Materials}\ }\textbf {\bibinfo {volume} {25}},\ \bibinfo
  {pages} {557} (\bibinfo {year} {2026})}\BibitemShut {NoStop}%
\bibitem [{\citenamefont {Randall}\ \emph {et~al.}(2021)\citenamefont
  {Randall}, \citenamefont {Fan}, \citenamefont {Reaney}, \citenamefont
  {Chen},\ and\ \citenamefont {Trolier-McKinstry}}]{AntiFerro}%
  \BibitemOpen
  \bibfield  {author} {\bibinfo {author} {\bibfnamefont {C.~A.}\ \bibnamefont
  {Randall}}, \bibinfo {author} {\bibfnamefont {Z.}~\bibnamefont {Fan}},
  \bibinfo {author} {\bibfnamefont {I.}~\bibnamefont {Reaney}}, \bibinfo
  {author} {\bibfnamefont {L.-Q.}\ \bibnamefont {Chen}},\ and\ \bibinfo
  {author} {\bibfnamefont {S.}~\bibnamefont {Trolier-McKinstry}},\ }\bibfield
  {title} {\bibinfo {title} {Antiferroelectrics: History, fundamentals, crystal
  chemistry, crystal structures, size effects, and applications},\ }\href
  {https://doi.org/https://doi.org/10.1111/jace.17834} {\bibfield  {journal}
  {\bibinfo  {journal} {Journal of the American Ceramic Society}\ }\textbf
  {\bibinfo {volume} {104}},\ \bibinfo {pages} {3775} (\bibinfo {year}
  {2021})}\BibitemShut {NoStop}%
\bibitem [{\citenamefont {Rabe}(2013)}]{AFE_Rabe}%
  \BibitemOpen
  \bibfield  {author} {\bibinfo {author} {\bibfnamefont {K.~M.}\ \bibnamefont
  {Rabe}},\ }\bibinfo {title} {Antiferroelectricity in oxides: A
  reexamination},\ in\ \href
  {https://doi.org/https://doi.org/10.1002/9783527654864.ch7} {\emph {\bibinfo
  {booktitle} {Functional Metal Oxides}}}\ (\bibinfo  {publisher} {John Wiley
  \& Sons, Ltd},\ \bibinfo {year} {2013})\ Chap.~\bibinfo {chapter} {7}, pp.\
  \bibinfo {pages} {221--244}\BibitemShut {NoStop}%
\bibitem [{\citenamefont {Si}\ \emph {et~al.}(2024)\citenamefont {Si},
  \citenamefont {Zhang}, \citenamefont {Liu}, \citenamefont {Das},
  \citenamefont {Xu}, \citenamefont {Burkovsky}, \citenamefont {Wei},\ and\
  \citenamefont {Chen}}]{SI2024101231}%
  \BibitemOpen
  \bibfield  {author} {\bibinfo {author} {\bibfnamefont {Y.}~\bibnamefont
  {Si}}, \bibinfo {author} {\bibfnamefont {T.}~\bibnamefont {Zhang}}, \bibinfo
  {author} {\bibfnamefont {C.}~\bibnamefont {Liu}}, \bibinfo {author}
  {\bibfnamefont {S.}~\bibnamefont {Das}}, \bibinfo {author} {\bibfnamefont
  {B.}~\bibnamefont {Xu}}, \bibinfo {author} {\bibfnamefont {R.~G.}\
  \bibnamefont {Burkovsky}}, \bibinfo {author} {\bibfnamefont {X.-K.}\
  \bibnamefont {Wei}},\ and\ \bibinfo {author} {\bibfnamefont {Z.}~\bibnamefont
  {Chen}},\ }\bibfield  {title} {\bibinfo {title} {Antiferroelectric oxide
  thin-films: Fundamentals, properties, and applications},\ }\href
  {https://doi.org/https://doi.org/10.1016/j.pmatsci.2023.101231} {\bibfield
  {journal} {\bibinfo  {journal} {Progress in Materials Science}\ }\textbf
  {\bibinfo {volume} {142}},\ \bibinfo {pages} {101231} (\bibinfo {year}
  {2024})}\BibitemShut {NoStop}%
\bibitem [{\citenamefont {Liu}\ \emph {et~al.}(2018)\citenamefont {Liu},
  \citenamefont {Lu}, \citenamefont {Ye}, \citenamefont {Wang}, \citenamefont
  {Dong}, \citenamefont {Withers},\ and\ \citenamefont {Liu}}]{AP_review}%
  \BibitemOpen
  \bibfield  {author} {\bibinfo {author} {\bibfnamefont {Z.}~\bibnamefont
  {Liu}}, \bibinfo {author} {\bibfnamefont {T.}~\bibnamefont {Lu}}, \bibinfo
  {author} {\bibfnamefont {J.}~\bibnamefont {Ye}}, \bibinfo {author}
  {\bibfnamefont {G.}~\bibnamefont {Wang}}, \bibinfo {author} {\bibfnamefont
  {X.}~\bibnamefont {Dong}}, \bibinfo {author} {\bibfnamefont {R.}~\bibnamefont
  {Withers}},\ and\ \bibinfo {author} {\bibfnamefont {Y.}~\bibnamefont {Liu}},\
  }\bibfield  {title} {\bibinfo {title} {Antiferroelectrics for energy storage
  applications: a review},\ }\href
  {https://doi.org/https://doi.org/10.1002/admt.201800111} {\bibfield
  {journal} {\bibinfo  {journal} {Advanced Materials Technologies}\ }\textbf
  {\bibinfo {volume} {3}},\ \bibinfo {pages} {1800111} (\bibinfo {year}
  {2018})}\BibitemShut {NoStop}%
\bibitem [{\citenamefont {Shirane}\ \emph {et~al.}(1951)\citenamefont
  {Shirane}, \citenamefont {Sawaguchi},\ and\ \citenamefont {Takagi}}]{PZO}%
  \BibitemOpen
  \bibfield  {author} {\bibinfo {author} {\bibfnamefont {G.}~\bibnamefont
  {Shirane}}, \bibinfo {author} {\bibfnamefont {E.}~\bibnamefont {Sawaguchi}},\
  and\ \bibinfo {author} {\bibfnamefont {Y.}~\bibnamefont {Takagi}},\
  }\bibfield  {title} {\bibinfo {title} {Dielectric properties of lead
  zirconate},\ }\href {https://doi.org/10.1103/PhysRev.84.476} {\bibfield
  {journal} {\bibinfo  {journal} {Phys. Rev.}\ }\textbf {\bibinfo {volume}
  {84}},\ \bibinfo {pages} {476} (\bibinfo {year} {1951})}\BibitemShut
  {NoStop}%
\bibitem [{\citenamefont {Tagantsev}\ \emph {et~al.}(2013)\citenamefont
  {Tagantsev}, \citenamefont {Vaideeswaran}, \citenamefont {Vakhrushev},
  \citenamefont {Filimonov}, \citenamefont {Burkovsky}, \citenamefont
  {Shaganov}, \citenamefont {Andronikova}, \citenamefont {Rudskoy},
  \citenamefont {Baron}, \citenamefont {Uchiyama}, \citenamefont {Chernyshov},
  \citenamefont {Bosak}, \citenamefont {Ujma}, \citenamefont {Roleder},
  \citenamefont {Majchrowski}, \citenamefont {Ko},\ and\ \citenamefont
  {Setter}}]{PZO_Anti}%
  \BibitemOpen
  \bibfield  {author} {\bibinfo {author} {\bibfnamefont {A.~K.}\ \bibnamefont
  {Tagantsev}}, \bibinfo {author} {\bibfnamefont {K.}~\bibnamefont
  {Vaideeswaran}}, \bibinfo {author} {\bibfnamefont {S.~B.}\ \bibnamefont
  {Vakhrushev}}, \bibinfo {author} {\bibfnamefont {A.~V.}\ \bibnamefont
  {Filimonov}}, \bibinfo {author} {\bibfnamefont {R.~G.}\ \bibnamefont
  {Burkovsky}}, \bibinfo {author} {\bibfnamefont {A.}~\bibnamefont {Shaganov}},
  \bibinfo {author} {\bibfnamefont {D.}~\bibnamefont {Andronikova}}, \bibinfo
  {author} {\bibfnamefont {A.~I.}\ \bibnamefont {Rudskoy}}, \bibinfo {author}
  {\bibfnamefont {A.~Q.~R.}\ \bibnamefont {Baron}}, \bibinfo {author}
  {\bibfnamefont {H.}~\bibnamefont {Uchiyama}}, \bibinfo {author}
  {\bibfnamefont {D.}~\bibnamefont {Chernyshov}}, \bibinfo {author}
  {\bibfnamefont {A.}~\bibnamefont {Bosak}}, \bibinfo {author} {\bibfnamefont
  {Z.}~\bibnamefont {Ujma}}, \bibinfo {author} {\bibfnamefont {K.}~\bibnamefont
  {Roleder}}, \bibinfo {author} {\bibfnamefont {A.}~\bibnamefont
  {Majchrowski}}, \bibinfo {author} {\bibfnamefont {J.-H.}\ \bibnamefont
  {Ko}},\ and\ \bibinfo {author} {\bibfnamefont {N.}~\bibnamefont {Setter}},\
  }\bibfield  {title} {\bibinfo {title} {The origin of antiferroelectricity in
  pbzro3},\ }\href {https://doi.org/10.1038/ncomms3229} {\bibfield  {journal}
  {\bibinfo  {journal} {Nature Communications}\ }\textbf {\bibinfo {volume}
  {4}},\ \bibinfo {pages} {2229} (\bibinfo {year} {2013})}\BibitemShut
  {NoStop}%
\bibitem [{\citenamefont {Zhao}\ \emph {et~al.}(2017)\citenamefont {Zhao},
  \citenamefont {Liu}, \citenamefont {Gao}, \citenamefont {Zhang},\ and\
  \citenamefont {Li}}]{AgNbO}%
  \BibitemOpen
  \bibfield  {author} {\bibinfo {author} {\bibfnamefont {L.}~\bibnamefont
  {Zhao}}, \bibinfo {author} {\bibfnamefont {Q.}~\bibnamefont {Liu}}, \bibinfo
  {author} {\bibfnamefont {J.}~\bibnamefont {Gao}}, \bibinfo {author}
  {\bibfnamefont {S.}~\bibnamefont {Zhang}},\ and\ \bibinfo {author}
  {\bibfnamefont {J.-F.}\ \bibnamefont {Li}},\ }\bibfield  {title} {\bibinfo
  {title} {Lead-free antiferroelectric silver niobate tantalate with high
  energy storage performance},\ }\href
  {https://doi.org/https://doi.org/10.1002/adma.201701824} {\bibfield
  {journal} {\bibinfo  {journal} {Advanced Materials}\ }\textbf {\bibinfo
  {volume} {29}},\ \bibinfo {pages} {1701824} (\bibinfo {year}
  {2017})}\BibitemShut {NoStop}%
\bibitem [{\citenamefont {Qi}\ \emph {et~al.}(2019)\citenamefont {Qi},
  \citenamefont {Zuo}, \citenamefont {Xie}, \citenamefont {Tian}, \citenamefont
  {Fu}, \citenamefont {Zhang},\ and\ \citenamefont {Zhang}}]{NaNbO}%
  \BibitemOpen
  \bibfield  {author} {\bibinfo {author} {\bibfnamefont {H.}~\bibnamefont
  {Qi}}, \bibinfo {author} {\bibfnamefont {R.}~\bibnamefont {Zuo}}, \bibinfo
  {author} {\bibfnamefont {A.}~\bibnamefont {Xie}}, \bibinfo {author}
  {\bibfnamefont {A.}~\bibnamefont {Tian}}, \bibinfo {author} {\bibfnamefont
  {J.}~\bibnamefont {Fu}}, \bibinfo {author} {\bibfnamefont {Y.}~\bibnamefont
  {Zhang}},\ and\ \bibinfo {author} {\bibfnamefont {S.}~\bibnamefont {Zhang}},\
  }\bibfield  {title} {\bibinfo {title} {Ultrahigh energy-storage density in
  nanbo3-based lead-free relaxor antiferroelectric ceramics with nanoscale
  domains},\ }\href {https://doi.org/https://doi.org/10.1002/adfm.201903877}
  {\bibfield  {journal} {\bibinfo  {journal} {Advanced Functional Materials}\
  }\textbf {\bibinfo {volume} {29}},\ \bibinfo {pages} {1903877} (\bibinfo
  {year} {2019})}\BibitemShut {NoStop}%
\bibitem [{\citenamefont {Alaerts}\ \emph {et~al.}(2026)\citenamefont
  {Alaerts}, \citenamefont {Schimpf}, \citenamefont {Li}, \citenamefont
  {Zheng}, \citenamefont {Banyas}, \citenamefont {Neaton}, \citenamefont
  {Griffin}, \citenamefont {Han}, \citenamefont {Martin},\ and\ \citenamefont
  {Hautier}}]{AFE_2026}%
  \BibitemOpen
  \bibfield  {author} {\bibinfo {author} {\bibfnamefont {L.}~\bibnamefont
  {Alaerts}}, \bibinfo {author} {\bibfnamefont {J.}~\bibnamefont {Schimpf}},
  \bibinfo {author} {\bibfnamefont {X.}~\bibnamefont {Li}}, \bibinfo {author}
  {\bibfnamefont {J.}~\bibnamefont {Zheng}}, \bibinfo {author} {\bibfnamefont
  {E.}~\bibnamefont {Banyas}}, \bibinfo {author} {\bibfnamefont {J.~B.}\
  \bibnamefont {Neaton}}, \bibinfo {author} {\bibfnamefont {S.~M.}\
  \bibnamefont {Griffin}}, \bibinfo {author} {\bibfnamefont {Y.}~\bibnamefont
  {Han}}, \bibinfo {author} {\bibfnamefont {L.~W.}\ \bibnamefont {Martin}},\
  and\ \bibinfo {author} {\bibfnamefont {G.}~\bibnamefont {Hautier}},\ }\href
  {https://arxiv.org/abs/2601.04916} {\bibinfo {title} {Discovery of a new
  weberite-type antiferroelectric: La3nbo7}} (\bibinfo {year} {2026}),\ \Eprint
  {https://arxiv.org/abs/2601.04916} {arXiv:2601.04916 [cond-mat.mtrl-sci]}
  \BibitemShut {NoStop}%
\bibitem [{\citenamefont {Zhang}\ \emph {et~al.}(2017)\citenamefont {Zhang},
  \citenamefont {Ma}, \citenamefont {Loh}, \citenamefont {Li}, \citenamefont
  {Walsh},\ and\ \citenamefont {Kong}}]{acsenergylett}%
  \BibitemOpen
  \bibfield  {author} {\bibinfo {author} {\bibfnamefont {W.-B.}\ \bibnamefont
  {Zhang}}, \bibinfo {author} {\bibfnamefont {X.-J.}\ \bibnamefont {Ma}},
  \bibinfo {author} {\bibfnamefont {A.}~\bibnamefont {Loh}}, \bibinfo {author}
  {\bibfnamefont {X.}~\bibnamefont {Li}}, \bibinfo {author} {\bibfnamefont
  {F.~C.}\ \bibnamefont {Walsh}},\ and\ \bibinfo {author} {\bibfnamefont
  {L.-B.}\ \bibnamefont {Kong}},\ }\bibfield  {title} {\bibinfo {title} {High
  volumetric energy density capacitors based on new electrode material
  lanthanum nitride},\ }\href {https://doi.org/10.1021/acsenergylett.6b00636}
  {\bibfield  {journal} {\bibinfo  {journal} {ACS Energy Letters}\ }\textbf
  {\bibinfo {volume} {2}},\ \bibinfo {pages} {336} (\bibinfo {year}
  {2017})}\BibitemShut {NoStop}%
\bibitem [{\citenamefont {Young}\ and\ \citenamefont
  {Ziegler}(1952)}]{jacs_1952}%
  \BibitemOpen
  \bibfield  {author} {\bibinfo {author} {\bibfnamefont {R.~A.}\ \bibnamefont
  {Young}}\ and\ \bibinfo {author} {\bibfnamefont {W.~T.}\ \bibnamefont
  {Ziegler}},\ }\bibfield  {title} {\bibinfo {title} {Crystal structure of
  lanthanum nitride1,2},\ }\href {https://doi.org/10.1021/ja01141a004}
  {\bibfield  {journal} {\bibinfo  {journal} {Journal of the American Chemical
  Society}\ }\textbf {\bibinfo {volume} {74}},\ \bibinfo {pages} {5251}
  (\bibinfo {year} {1952})}\BibitemShut {NoStop}%
\bibitem [{\citenamefont {Krause}\ \emph {et~al.}(2018)\citenamefont {Krause},
  \citenamefont {Kuznetsov}, \citenamefont {Yakshin}, \citenamefont
  {Ibrahimkutty}, \citenamefont {Baumbach},\ and\ \citenamefont
  {Bijkerk}}]{Krause_nb5222}%
  \BibitemOpen
  \bibfield  {author} {\bibinfo {author} {\bibfnamefont {B.}~\bibnamefont
  {Krause}}, \bibinfo {author} {\bibfnamefont {D.~S.}\ \bibnamefont
  {Kuznetsov}}, \bibinfo {author} {\bibfnamefont {A.~E.}\ \bibnamefont
  {Yakshin}}, \bibinfo {author} {\bibfnamefont {S.}~\bibnamefont
  {Ibrahimkutty}}, \bibinfo {author} {\bibfnamefont {T.}~\bibnamefont
  {Baumbach}},\ and\ \bibinfo {author} {\bibfnamefont {F.}~\bibnamefont
  {Bijkerk}},\ }\bibfield  {title} {\bibinfo {title} {{{\it In situ} and
  real-time monitoring of structure formation during non-reactive sputter
  deposition of lanthanum and reactive sputter deposition of lanthanum
  nitride}},\ }\href {https://doi.org/10.1107/S1600576718007367} {\bibfield
  {journal} {\bibinfo  {journal} {Journal of Applied Crystallography}\ }\textbf
  {\bibinfo {volume} {51}},\ \bibinfo {pages} {1013} (\bibinfo {year}
  {2018})}\BibitemShut {NoStop}%
\bibitem [{\citenamefont {Schneider}\ \emph {et~al.}(2012)\citenamefont
  {Schneider}, \citenamefont {Baumann}, \citenamefont {Salamat},\ and\
  \citenamefont {Schnick}}]{LaN_press}%
  \BibitemOpen
  \bibfield  {author} {\bibinfo {author} {\bibfnamefont {S.~B.}\ \bibnamefont
  {Schneider}}, \bibinfo {author} {\bibfnamefont {D.}~\bibnamefont {Baumann}},
  \bibinfo {author} {\bibfnamefont {A.}~\bibnamefont {Salamat}},\ and\ \bibinfo
  {author} {\bibfnamefont {W.}~\bibnamefont {Schnick}},\ }\bibfield  {title}
  {\bibinfo {title} {Reversible high-pressure phase transition in lan},\ }\href
  {https://doi.org/10.1063/1.4709392} {\bibfield  {journal} {\bibinfo
  {journal} {Journal of Applied Physics}\ }\textbf {\bibinfo {volume} {111}},\
  \bibinfo {pages} {093503} (\bibinfo {year} {2012})}\BibitemShut {NoStop}%
\bibitem [{\citenamefont {Chen}\ \emph {et~al.}(2021)\citenamefont {Chen},
  \citenamefont {Lin}, \citenamefont {Maciejko},\ and\ \citenamefont
  {Chen}}]{CHEN2021110779}%
  \BibitemOpen
  \bibfield  {author} {\bibinfo {author} {\bibfnamefont {W.-C.}\ \bibnamefont
  {Chen}}, \bibinfo {author} {\bibfnamefont {C.-M.}\ \bibnamefont {Lin}},
  \bibinfo {author} {\bibfnamefont {J.}~\bibnamefont {Maciejko}},\ and\
  \bibinfo {author} {\bibfnamefont {C.-C.}\ \bibnamefont {Chen}},\ }\bibfield
  {title} {\bibinfo {title} {Lan structural and topological transitions driven
  by temperature and pressure},\ }\href
  {https://doi.org/https://doi.org/10.1016/j.commatsci.2021.110779} {\bibfield
  {journal} {\bibinfo  {journal} {Computational Materials Science}\ }\textbf
  {\bibinfo {volume} {200}},\ \bibinfo {pages} {110779} (\bibinfo {year}
  {2021})}\BibitemShut {NoStop}%
\bibitem [{\citenamefont {Rowberg}\ \emph {et~al.}(2021)\citenamefont
  {Rowberg}, \citenamefont {Mu}, \citenamefont {Swift},\ and\ \citenamefont
  {Van~de Walle}}]{PhysRevMaterials.5.094602}%
  \BibitemOpen
  \bibfield  {author} {\bibinfo {author} {\bibfnamefont {A.~J.~E.}\
  \bibnamefont {Rowberg}}, \bibinfo {author} {\bibfnamefont {S.}~\bibnamefont
  {Mu}}, \bibinfo {author} {\bibfnamefont {M.~W.}\ \bibnamefont {Swift}},\ and\
  \bibinfo {author} {\bibfnamefont {C.~G.}\ \bibnamefont {Van~de Walle}},\
  }\bibfield  {title} {\bibinfo {title} {Structural, electronic, and
  polarization properties of yn and lan},\ }\href
  {https://doi.org/10.1103/PhysRevMaterials.5.094602} {\bibfield  {journal}
  {\bibinfo  {journal} {Phys. Rev. Mater.}\ }\textbf {\bibinfo {volume} {5}},\
  \bibinfo {pages} {094602} (\bibinfo {year} {2021})}\BibitemShut {NoStop}%
\bibitem [{\citenamefont {Chen}\ \emph {et~al.}(2025)\citenamefont {Chen},
  \citenamefont {Wang}, \citenamefont {Tejerina}, \citenamefont {Yazawa},
  \citenamefont {Zakutayev}, \citenamefont {Paillard},\ and\ \citenamefont
  {Bellaiche}}]{Peng}%
  \BibitemOpen
  \bibfield  {author} {\bibinfo {author} {\bibfnamefont {P.}~\bibnamefont
  {Chen}}, \bibinfo {author} {\bibfnamefont {D.}~\bibnamefont {Wang}}, \bibinfo
  {author} {\bibfnamefont {A.~M.}\ \bibnamefont {Tejerina}}, \bibinfo {author}
  {\bibfnamefont {K.}~\bibnamefont {Yazawa}}, \bibinfo {author} {\bibfnamefont
  {A.}~\bibnamefont {Zakutayev}}, \bibinfo {author} {\bibfnamefont
  {C.}~\bibnamefont {Paillard}},\ and\ \bibinfo {author} {\bibfnamefont
  {L.}~\bibnamefont {Bellaiche}},\ }\bibfield  {title} {\bibinfo {title}
  {Towards a deeper fundamental understanding of (al,sc)n ferroelectric
  nitrides},\ }\href {https://doi.org/10.1103/9nv5-ryqr} {\bibfield  {journal}
  {\bibinfo  {journal} {Phys. Rev. Mater.}\ }\textbf {\bibinfo {volume} {9}},\
  \bibinfo {pages} {124418} (\bibinfo {year} {2025})}\BibitemShut {NoStop}%
\bibitem [{SM()}]{SM}%
  \BibitemOpen
  \href@noop {} {\bibinfo {title} {See supplemental material at [url will be
  inserted by publisher] for technical calculations
  details~\cite{VASP1,VASP2,PAW,PBE,hse1,hse2,phonopy1,phonopy2,tdep1,tdep2,stdep},
  volume and born effective charges changes under pressure, as well as the dft
  potential energy surface.}}\BibitemShut {Stop}%
\bibitem [{\citenamefont {Ding}\ \emph {et~al.}(2022)\citenamefont {Ding},
  \citenamefont {Yuan}, \citenamefont {Cogollo-Olivo}, \citenamefont {Wang},
  \citenamefont {Wang},\ and\ \citenamefont {Sun}}]{Ding2022}%
  \BibitemOpen
  \bibfield  {author} {\bibinfo {author} {\bibfnamefont {C.}~\bibnamefont
  {Ding}}, \bibinfo {author} {\bibfnamefont {J.}~\bibnamefont {Yuan}}, \bibinfo
  {author} {\bibfnamefont {B.~H.}\ \bibnamefont {Cogollo-Olivo}}, \bibinfo
  {author} {\bibfnamefont {Y.}~\bibnamefont {Wang}}, \bibinfo {author}
  {\bibfnamefont {X.}~\bibnamefont {Wang}},\ and\ \bibinfo {author}
  {\bibfnamefont {J.}~\bibnamefont {Sun}},\ }\bibfield  {title} {\bibinfo
  {title} {Pressure-induced ferroelectric and anti-ferroelectric phase
  transitions in lan},\ }\href {https://doi.org/10.1007/s11433-022-1980-4}
  {\bibfield  {journal} {\bibinfo  {journal} {Science China Physics, Mechanics
  {\&} Astronomy}\ }\textbf {\bibinfo {volume} {66}},\ \bibinfo {pages}
  {228211} (\bibinfo {year} {2022})}\BibitemShut {NoStop}%
\bibitem [{\citenamefont {Aramberri}\ and\ \citenamefont
  {{\'I}{\~{n}}iguez}(2020)}]{Aramberri2020}%
  \BibitemOpen
  \bibfield  {author} {\bibinfo {author} {\bibfnamefont {H.}~\bibnamefont
  {Aramberri}}\ and\ \bibinfo {author} {\bibfnamefont {J.}~\bibnamefont
  {{\'I}{\~{n}}iguez}},\ }\bibfield  {title} {\bibinfo {title}
  {Antiferroelectricity in a family of pyroxene-like oxides with rich
  polymorphism},\ }\href {https://doi.org/10.1038/s43246-020-00051-9}
  {\bibfield  {journal} {\bibinfo  {journal} {Communications Materials}\
  }\textbf {\bibinfo {volume} {1}},\ \bibinfo {pages} {52} (\bibinfo {year}
  {2020})}\BibitemShut {NoStop}%
\bibitem [{\citenamefont {King-Smith}\ and\ \citenamefont
  {Vanderbilt}(1993)}]{Born_P1}%
  \BibitemOpen
  \bibfield  {author} {\bibinfo {author} {\bibfnamefont {R.~D.}\ \bibnamefont
  {King-Smith}}\ and\ \bibinfo {author} {\bibfnamefont {D.}~\bibnamefont
  {Vanderbilt}},\ }\bibfield  {title} {\bibinfo {title} {Theory of polarization
  of crystalline solids},\ }\href {https://doi.org/10.1103/PhysRevB.47.1651}
  {\bibfield  {journal} {\bibinfo  {journal} {Phys. Rev. B}\ }\textbf {\bibinfo
  {volume} {47}},\ \bibinfo {pages} {1651} (\bibinfo {year}
  {1993})}\BibitemShut {NoStop}%
\bibitem [{\citenamefont {Resta}(1994)}]{Resta}%
  \BibitemOpen
  \bibfield  {author} {\bibinfo {author} {\bibfnamefont {R.}~\bibnamefont
  {Resta}},\ }\bibfield  {title} {\bibinfo {title} {Macroscopic polarization in
  crystalline dielectrics: the geometric phase approach},\ }\href
  {https://doi.org/10.1103/RevModPhys.66.899} {\bibfield  {journal} {\bibinfo
  {journal} {Rev. Mod. Phys.}\ }\textbf {\bibinfo {volume} {66}},\ \bibinfo
  {pages} {899} (\bibinfo {year} {1994})}\BibitemShut {NoStop}%
\bibitem [{\citenamefont {Zheng}\ \emph {et~al.}(2026)\citenamefont {Zheng},
  \citenamefont {Paillard}, \citenamefont {Wang}, \citenamefont {Chen},
  \citenamefont {Zhao}, \citenamefont {Xie},\ and\ \citenamefont
  {Bellaiche}}]{Zheng2026}%
  \BibitemOpen
  \bibfield  {author} {\bibinfo {author} {\bibfnamefont {X.}~\bibnamefont
  {Zheng}}, \bibinfo {author} {\bibfnamefont {C.}~\bibnamefont {Paillard}},
  \bibinfo {author} {\bibfnamefont {D.}~\bibnamefont {Wang}}, \bibinfo {author}
  {\bibfnamefont {P.}~\bibnamefont {Chen}}, \bibinfo {author} {\bibfnamefont
  {H.~J.}\ \bibnamefont {Zhao}}, \bibinfo {author} {\bibfnamefont
  {Y.}~\bibnamefont {Xie}},\ and\ \bibinfo {author} {\bibfnamefont
  {L.}~\bibnamefont {Bellaiche}},\ }\bibfield  {title} {\bibinfo {title}
  {Domain-wall-mediated polarization switching in ferroelectric alscn: Strain
  relief and field-dependent dynamics},\ }\bibfield  {journal} {\bibinfo
  {journal} {Physical Review Letters}\ }\href
  {https://doi.org/10.1103/s8qs-nnzg} {10.1103/s8qs-nnzg} (\bibinfo {year}
  {2026})\BibitemShut {NoStop}%
\bibitem [{\citenamefont {Yazawa}\ \emph {et~al.}(2023)\citenamefont {Yazawa},
  \citenamefont {Hayden}, \citenamefont {Maria}, \citenamefont {Zhu},
  \citenamefont {Trolier-McKinstry}, \citenamefont {Zakutayev},\ and\
  \citenamefont {Brennecka}}]{Yazawa2023}%
  \BibitemOpen
  \bibfield  {author} {\bibinfo {author} {\bibfnamefont {K.}~\bibnamefont
  {Yazawa}}, \bibinfo {author} {\bibfnamefont {J.}~\bibnamefont {Hayden}},
  \bibinfo {author} {\bibfnamefont {J.-P.}\ \bibnamefont {Maria}}, \bibinfo
  {author} {\bibfnamefont {W.}~\bibnamefont {Zhu}}, \bibinfo {author}
  {\bibfnamefont {S.}~\bibnamefont {Trolier-McKinstry}}, \bibinfo {author}
  {\bibfnamefont {A.}~\bibnamefont {Zakutayev}},\ and\ \bibinfo {author}
  {\bibfnamefont {G.~L.}\ \bibnamefont {Brennecka}},\ }\bibfield  {title}
  {\bibinfo {title} {Anomalously abrupt switching of wurtzite-structured
  ferroelectrics: simultaneous non-linear nucleation and growth model},\ }\href
  {https://doi.org/10.1039/D3MH00365E} {\bibfield  {journal} {\bibinfo
  {journal} {Materials Horizons}\ }\textbf {\bibinfo {volume} {10}},\ \bibinfo
  {pages} {2936} (\bibinfo {year} {2023})}\BibitemShut {NoStop}%
\bibitem [{\citenamefont {Hwang}\ \emph {et~al.}(2024)\citenamefont {Hwang},
  \citenamefont {Aigner}, \citenamefont {Metzger}, \citenamefont {Kummel},\
  and\ \citenamefont {Cho}}]{AlN_height}%
  \BibitemOpen
  \bibfield  {author} {\bibinfo {author} {\bibfnamefont {T.}~\bibnamefont
  {Hwang}}, \bibinfo {author} {\bibfnamefont {W.}~\bibnamefont {Aigner}},
  \bibinfo {author} {\bibfnamefont {T.}~\bibnamefont {Metzger}}, \bibinfo
  {author} {\bibfnamefont {A.~C.}\ \bibnamefont {Kummel}},\ and\ \bibinfo
  {author} {\bibfnamefont {K.}~\bibnamefont {Cho}},\ }\bibfield  {title}
  {\bibinfo {title} {First-principles understanding on the formation of
  inversion domain boundaries of wurtzite aln, alscn, and gan},\ }\href
  {https://doi.org/10.1021/acsaelm.4c00097} {\bibfield  {journal} {\bibinfo
  {journal} {ACS Applied Electronic Materials}\ }\textbf {\bibinfo {volume}
  {6}},\ \bibinfo {pages} {3257} (\bibinfo {year} {2024})}\BibitemShut
  {NoStop}%
\bibitem [{\citenamefont {Bellaiche}\ and\ \citenamefont
  {\'I\~niguez}(2013)}]{Laurent}%
  \BibitemOpen
  \bibfield  {author} {\bibinfo {author} {\bibfnamefont {L.}~\bibnamefont
  {Bellaiche}}\ and\ \bibinfo {author} {\bibfnamefont {J.}~\bibnamefont
  {\'I\~niguez}},\ }\bibfield  {title} {\bibinfo {title} {Universal
  collaborative couplings between oxygen-octahedral rotations and
  antiferroelectric distortions in perovskites},\ }\href
  {https://doi.org/10.1103/PhysRevB.88.014104} {\bibfield  {journal} {\bibinfo
  {journal} {Phys. Rev. B}\ }\textbf {\bibinfo {volume} {88}},\ \bibinfo
  {pages} {014104} (\bibinfo {year} {2013})}\BibitemShut {NoStop}%
\bibitem [{\citenamefont {Patel}\ \emph {et~al.}(2016)\citenamefont {Patel},
  \citenamefont {Prosandeev}, \citenamefont {Yang}, \citenamefont {Xu},
  \citenamefont {\'I\~niguez},\ and\ \citenamefont {Bellaiche}}]{Kinnary}%
  \BibitemOpen
  \bibfield  {author} {\bibinfo {author} {\bibfnamefont {K.}~\bibnamefont
  {Patel}}, \bibinfo {author} {\bibfnamefont {S.}~\bibnamefont {Prosandeev}},
  \bibinfo {author} {\bibfnamefont {Y.}~\bibnamefont {Yang}}, \bibinfo {author}
  {\bibfnamefont {B.}~\bibnamefont {Xu}}, \bibinfo {author} {\bibfnamefont
  {J.}~\bibnamefont {\'I\~niguez}},\ and\ \bibinfo {author} {\bibfnamefont
  {L.}~\bibnamefont {Bellaiche}},\ }\bibfield  {title} {\bibinfo {title}
  {Atomistic mechanism leading to complex antiferroelectric and incommensurate
  perovskites},\ }\href {https://doi.org/10.1103/PhysRevB.94.054107} {\bibfield
   {journal} {\bibinfo  {journal} {Phys. Rev. B}\ }\textbf {\bibinfo {volume}
  {94}},\ \bibinfo {pages} {054107} (\bibinfo {year} {2016})}\BibitemShut
  {NoStop}%
\bibitem [{\citenamefont {Kresse}\ and\ \citenamefont {Hafner}(1993)}]{VASP1}%
  \BibitemOpen
  \bibfield  {author} {\bibinfo {author} {\bibfnamefont {G.}~\bibnamefont
  {Kresse}}\ and\ \bibinfo {author} {\bibfnamefont {J.}~\bibnamefont
  {Hafner}},\ }\bibfield  {title} {\bibinfo {title} {Ab initio molecular
  dynamics for liquid metals},\ }\href
  {https://doi.org/10.1103/PhysRevB.47.558} {\bibfield  {journal} {\bibinfo
  {journal} {Phys. Rev. B}\ }\textbf {\bibinfo {volume} {47}},\ \bibinfo
  {pages} {558} (\bibinfo {year} {1993})}\BibitemShut {NoStop}%
\bibitem [{\citenamefont {Kresse}\ and\ \citenamefont
  {Furthm\"uller}(1996)}]{VASP2}%
  \BibitemOpen
  \bibfield  {author} {\bibinfo {author} {\bibfnamefont {G.}~\bibnamefont
  {Kresse}}\ and\ \bibinfo {author} {\bibfnamefont {J.}~\bibnamefont
  {Furthm\"uller}},\ }\bibfield  {title} {\bibinfo {title} {Efficient iterative
  schemes for ab initio total-energy calculations using a plane-wave basis
  set},\ }\href {https://doi.org/10.1103/PhysRevB.54.11169} {\bibfield
  {journal} {\bibinfo  {journal} {Phys. Rev. B}\ }\textbf {\bibinfo {volume}
  {54}},\ \bibinfo {pages} {11169} (\bibinfo {year} {1996})}\BibitemShut
  {NoStop}%
\bibitem [{\citenamefont {Bl\"ochl}(1994)}]{PAW}%
  \BibitemOpen
  \bibfield  {author} {\bibinfo {author} {\bibfnamefont {P.~E.}\ \bibnamefont
  {Bl\"ochl}},\ }\bibfield  {title} {\bibinfo {title} {Projector augmented-wave
  method},\ }\href {https://doi.org/10.1103/PhysRevB.50.17953} {\bibfield
  {journal} {\bibinfo  {journal} {Phys. Rev. B}\ }\textbf {\bibinfo {volume}
  {50}},\ \bibinfo {pages} {17953} (\bibinfo {year} {1994})}\BibitemShut
  {NoStop}%
\bibitem [{\citenamefont {Perdew}\ \emph {et~al.}(1996)\citenamefont {Perdew},
  \citenamefont {Burke},\ and\ \citenamefont {Ernzerhof}}]{PBE}%
  \BibitemOpen
  \bibfield  {author} {\bibinfo {author} {\bibfnamefont {J.~P.}\ \bibnamefont
  {Perdew}}, \bibinfo {author} {\bibfnamefont {K.}~\bibnamefont {Burke}},\ and\
  \bibinfo {author} {\bibfnamefont {M.}~\bibnamefont {Ernzerhof}},\ }\bibfield
  {title} {\bibinfo {title} {Generalized gradient approximation made simple},\
  }\href {https://doi.org/10.1103/PhysRevLett.77.3865} {\bibfield  {journal}
  {\bibinfo  {journal} {Phys. Rev. Lett.}\ }\textbf {\bibinfo {volume} {77}},\
  \bibinfo {pages} {3865} (\bibinfo {year} {1996})}\BibitemShut {NoStop}%
\bibitem [{\citenamefont {Heyd}\ \emph {et~al.}(2003)\citenamefont {Heyd},
  \citenamefont {Scuseria},\ and\ \citenamefont {Ernzerhof}}]{hse1}%
  \BibitemOpen
  \bibfield  {author} {\bibinfo {author} {\bibfnamefont {J.}~\bibnamefont
  {Heyd}}, \bibinfo {author} {\bibfnamefont {G.~E.}\ \bibnamefont {Scuseria}},\
  and\ \bibinfo {author} {\bibfnamefont {M.}~\bibnamefont {Ernzerhof}},\
  }\bibfield  {title} {\bibinfo {title} {Hybrid functionals based on a screened
  coulomb potential},\ }\href {https://doi.org/10.1063/1.1564060} {\bibfield
  {journal} {\bibinfo  {journal} {The Journal of Chemical Physics}\ }\textbf
  {\bibinfo {volume} {118}},\ \bibinfo {pages} {8207} (\bibinfo {year}
  {2003})}\BibitemShut {NoStop}%
\bibitem [{\citenamefont {Krukau}\ \emph {et~al.}(2006)\citenamefont {Krukau},
  \citenamefont {Vydrov}, \citenamefont {Izmaylov},\ and\ \citenamefont
  {Scuseria}}]{hse2}%
  \BibitemOpen
  \bibfield  {author} {\bibinfo {author} {\bibfnamefont {A.~V.}\ \bibnamefont
  {Krukau}}, \bibinfo {author} {\bibfnamefont {O.~A.}\ \bibnamefont {Vydrov}},
  \bibinfo {author} {\bibfnamefont {A.~F.}\ \bibnamefont {Izmaylov}},\ and\
  \bibinfo {author} {\bibfnamefont {G.~E.}\ \bibnamefont {Scuseria}},\
  }\bibfield  {title} {\bibinfo {title} {Influence of the exchange screening
  parameter on the performance of screened hybrid functionals},\ }\href
  {https://doi.org/10.1063/1.2404663} {\bibfield  {journal} {\bibinfo
  {journal} {The Journal of Chemical Physics}\ }\textbf {\bibinfo {volume}
  {125}},\ \bibinfo {pages} {224106} (\bibinfo {year} {2006})}\BibitemShut
  {NoStop}%
\bibitem [{\citenamefont {Togo}\ \emph {et~al.}(2023)\citenamefont {Togo},
  \citenamefont {Chaput}, \citenamefont {Tadano},\ and\ \citenamefont
  {Tanaka}}]{phonopy1}%
  \BibitemOpen
  \bibfield  {author} {\bibinfo {author} {\bibfnamefont {A.}~\bibnamefont
  {Togo}}, \bibinfo {author} {\bibfnamefont {L.}~\bibnamefont {Chaput}},
  \bibinfo {author} {\bibfnamefont {T.}~\bibnamefont {Tadano}},\ and\ \bibinfo
  {author} {\bibfnamefont {I.}~\bibnamefont {Tanaka}},\ }\bibfield  {title}
  {\bibinfo {title} {Implementation strategies in phonopy and phono3py},\
  }\href {https://doi.org/10.1088/1361-648X/acd831} {\bibfield  {journal}
  {\bibinfo  {journal} {J. Phys. Condens. Matter}\ }\textbf {\bibinfo {volume}
  {35}},\ \bibinfo {pages} {353001} (\bibinfo {year} {2023})}\BibitemShut
  {NoStop}%
\bibitem [{\citenamefont {Togo}(2023)}]{phonopy2}%
  \BibitemOpen
  \bibfield  {author} {\bibinfo {author} {\bibfnamefont {A.}~\bibnamefont
  {Togo}},\ }\bibfield  {title} {\bibinfo {title} {First-principles phonon
  calculations with phonopy and phono3py},\ }\href
  {https://doi.org/10.7566/JPSJ.92.012001} {\bibfield  {journal} {\bibinfo
  {journal} {J. Phys. Soc. Jpn.}\ }\textbf {\bibinfo {volume} {92}},\ \bibinfo
  {pages} {012001} (\bibinfo {year} {2023})}\BibitemShut {NoStop}%
\bibitem [{\citenamefont {Hellman}\ \emph {et~al.}(2011)\citenamefont
  {Hellman}, \citenamefont {Abrikosov},\ and\ \citenamefont {Simak}}]{tdep1}%
  \BibitemOpen
  \bibfield  {author} {\bibinfo {author} {\bibfnamefont {O.}~\bibnamefont
  {Hellman}}, \bibinfo {author} {\bibfnamefont {I.~A.}\ \bibnamefont
  {Abrikosov}},\ and\ \bibinfo {author} {\bibfnamefont {S.~I.}\ \bibnamefont
  {Simak}},\ }\bibfield  {title} {\bibinfo {title} {Lattice dynamics of
  anharmonic solids from first principles},\ }\href
  {https://doi.org/10.1103/PhysRevB.84.180301} {\bibfield  {journal} {\bibinfo
  {journal} {Phys. Rev. B}\ }\textbf {\bibinfo {volume} {84}},\ \bibinfo
  {pages} {180301} (\bibinfo {year} {2011})}\BibitemShut {NoStop}%
\bibitem [{\citenamefont {Hellman}\ and\ \citenamefont
  {Abrikosov}(2013)}]{tdep2}%
  \BibitemOpen
  \bibfield  {author} {\bibinfo {author} {\bibfnamefont {O.}~\bibnamefont
  {Hellman}}\ and\ \bibinfo {author} {\bibfnamefont {I.~A.}\ \bibnamefont
  {Abrikosov}},\ }\bibfield  {title} {\bibinfo {title} {Temperature-dependent
  effective third-order interatomic force constants from first principles},\
  }\href {https://doi.org/10.1103/PhysRevB.88.144301} {\bibfield  {journal}
  {\bibinfo  {journal} {Phys. Rev. B}\ }\textbf {\bibinfo {volume} {88}},\
  \bibinfo {pages} {144301} (\bibinfo {year} {2013})}\BibitemShut {NoStop}%
\bibitem [{\citenamefont {Shulumba}\ \emph {et~al.}(2017)\citenamefont
  {Shulumba}, \citenamefont {Hellman},\ and\ \citenamefont {Minnich}}]{stdep}%
  \BibitemOpen
  \bibfield  {author} {\bibinfo {author} {\bibfnamefont {N.}~\bibnamefont
  {Shulumba}}, \bibinfo {author} {\bibfnamefont {O.}~\bibnamefont {Hellman}},\
  and\ \bibinfo {author} {\bibfnamefont {A.~J.}\ \bibnamefont {Minnich}},\
  }\bibfield  {title} {\bibinfo {title} {Intrinsic localized mode and low
  thermal conductivity of pbse},\ }\href
  {https://doi.org/10.1103/PhysRevB.95.014302} {\bibfield  {journal} {\bibinfo
  {journal} {Phys. Rev. B}\ }\textbf {\bibinfo {volume} {95}},\ \bibinfo
  {pages} {014302} (\bibinfo {year} {2017})}\BibitemShut {NoStop}%
\end{thebibliography}%


\providecommand{\noopsort}[1]{}\providecommand{\singleletter}[1]{#1}%
\begin{thebibliography}{12}%
\makeatletter
\providecommand \@ifxundefined [1]{%
 \@ifx{#1\undefined}
}%
\providecommand \@ifnum [1]{%
 \ifnum #1\expandafter \@firstoftwo
 \else \expandafter \@secondoftwo
 \fi
}%
\providecommand \@ifx [1]{%
 \ifx #1\expandafter \@firstoftwo
 \else \expandafter \@secondoftwo
 \fi
}%
\providecommand \natexlab [1]{#1}%
\providecommand \enquote  [1]{``#1''}%
\providecommand \bibnamefont  [1]{#1}%
\providecommand \bibfnamefont [1]{#1}%
\providecommand \citenamefont [1]{#1}%
\providecommand \href@noop [0]{\@secondoftwo}%
\providecommand \href [0]{\begingroup \@sanitize@url \@href}%
\providecommand \@href[1]{\@@startlink{#1}\@@href}%
\providecommand \@@href[1]{\endgroup#1\@@endlink}%
\providecommand \@sanitize@url [0]{\catcode `\\12\catcode `\$12\catcode
  `\&12\catcode `\#12\catcode `\^12\catcode `\_12\catcode `\%12\relax}%
\providecommand \@@startlink[1]{}%
\providecommand \@@endlink[0]{}%
\providecommand \url  [0]{\begingroup\@sanitize@url \@url }%
\providecommand \@url [1]{\endgroup\@href {#1}{\urlprefix }}%
\providecommand \urlprefix  [0]{URL }%
\providecommand \Eprint [0]{\href }%
\providecommand \doibase [0]{https://doi.org/}%
\providecommand \selectlanguage [0]{\@gobble}%
\providecommand \bibinfo  [0]{\@secondoftwo}%
\providecommand \bibfield  [0]{\@secondoftwo}%
\providecommand \translation [1]{[#1]}%
\providecommand \BibitemOpen [0]{}%
\providecommand \bibitemStop [0]{}%
\providecommand \bibitemNoStop [0]{.\EOS\space}%
\providecommand \EOS [0]{\spacefactor3000\relax}%
\providecommand \BibitemShut  [1]{\csname bibitem#1\endcsname}%
\let\auto@bib@innerbib\@empty
\bibitem [{\citenamefont {Kresse}\ and\ \citenamefont {Hafner}(1993)}]{VASP1}%
  \BibitemOpen
  \bibfield  {author} {\bibinfo {author} {\bibfnamefont {G.}~\bibnamefont
  {Kresse}}\ and\ \bibinfo {author} {\bibfnamefont {J.}~\bibnamefont
  {Hafner}},\ }\bibfield  {title} {\bibinfo {title} {Ab initio molecular
  dynamics for liquid metals},\ }\href
  {https://doi.org/10.1103/PhysRevB.47.558} {\bibfield  {journal} {\bibinfo
  {journal} {Phys. Rev. B}\ }\textbf {\bibinfo {volume} {47}},\ \bibinfo
  {pages} {558} (\bibinfo {year} {1993})}\BibitemShut {NoStop}%
\bibitem [{\citenamefont {Kresse}\ and\ \citenamefont
  {Furthm\"uller}(1996)}]{VASP2}%
  \BibitemOpen
  \bibfield  {author} {\bibinfo {author} {\bibfnamefont {G.}~\bibnamefont
  {Kresse}}\ and\ \bibinfo {author} {\bibfnamefont {J.}~\bibnamefont
  {Furthm\"uller}},\ }\bibfield  {title} {\bibinfo {title} {Efficient iterative
  schemes for ab initio total-energy calculations using a plane-wave basis
  set},\ }\href {https://doi.org/10.1103/PhysRevB.54.11169} {\bibfield
  {journal} {\bibinfo  {journal} {Phys. Rev. B}\ }\textbf {\bibinfo {volume}
  {54}},\ \bibinfo {pages} {11169} (\bibinfo {year} {1996})}\BibitemShut
  {NoStop}%
\bibitem [{\citenamefont {Bl\"ochl}(1994)}]{PAW}%
  \BibitemOpen
  \bibfield  {author} {\bibinfo {author} {\bibfnamefont {P.~E.}\ \bibnamefont
  {Bl\"ochl}},\ }\bibfield  {title} {\bibinfo {title} {Projector augmented-wave
  method},\ }\href {https://doi.org/10.1103/PhysRevB.50.17953} {\bibfield
  {journal} {\bibinfo  {journal} {Phys. Rev. B}\ }\textbf {\bibinfo {volume}
  {50}},\ \bibinfo {pages} {17953} (\bibinfo {year} {1994})}\BibitemShut
  {NoStop}%
\bibitem [{\citenamefont {Perdew}\ \emph {et~al.}(1996)\citenamefont {Perdew},
  \citenamefont {Burke},\ and\ \citenamefont {Ernzerhof}}]{PBE}%
  \BibitemOpen
  \bibfield  {author} {\bibinfo {author} {\bibfnamefont {J.~P.}\ \bibnamefont
  {Perdew}}, \bibinfo {author} {\bibfnamefont {K.}~\bibnamefont {Burke}},\ and\
  \bibinfo {author} {\bibfnamefont {M.}~\bibnamefont {Ernzerhof}},\ }\bibfield
  {title} {\bibinfo {title} {Generalized gradient approximation made simple},\
  }\href {https://doi.org/10.1103/PhysRevLett.77.3865} {\bibfield  {journal}
  {\bibinfo  {journal} {Phys. Rev. Lett.}\ }\textbf {\bibinfo {volume} {77}},\
  \bibinfo {pages} {3865} (\bibinfo {year} {1996})}\BibitemShut {NoStop}%
\bibitem [{\citenamefont {Heyd}\ \emph {et~al.}(2003)\citenamefont {Heyd},
  \citenamefont {Scuseria},\ and\ \citenamefont {Ernzerhof}}]{hse1}%
  \BibitemOpen
  \bibfield  {author} {\bibinfo {author} {\bibfnamefont {J.}~\bibnamefont
  {Heyd}}, \bibinfo {author} {\bibfnamefont {G.~E.}\ \bibnamefont {Scuseria}},\
  and\ \bibinfo {author} {\bibfnamefont {M.}~\bibnamefont {Ernzerhof}},\
  }\bibfield  {title} {\bibinfo {title} {Hybrid functionals based on a screened
  coulomb potential},\ }\href {https://doi.org/10.1063/1.1564060} {\bibfield
  {journal} {\bibinfo  {journal} {The Journal of Chemical Physics}\ }\textbf
  {\bibinfo {volume} {118}},\ \bibinfo {pages} {8207} (\bibinfo {year}
  {2003})}\BibitemShut {NoStop}%
\bibitem [{\citenamefont {Krukau}\ \emph {et~al.}(2006)\citenamefont {Krukau},
  \citenamefont {Vydrov}, \citenamefont {Izmaylov},\ and\ \citenamefont
  {Scuseria}}]{hse2}%
  \BibitemOpen
  \bibfield  {author} {\bibinfo {author} {\bibfnamefont {A.~V.}\ \bibnamefont
  {Krukau}}, \bibinfo {author} {\bibfnamefont {O.~A.}\ \bibnamefont {Vydrov}},
  \bibinfo {author} {\bibfnamefont {A.~F.}\ \bibnamefont {Izmaylov}},\ and\
  \bibinfo {author} {\bibfnamefont {G.~E.}\ \bibnamefont {Scuseria}},\
  }\bibfield  {title} {\bibinfo {title} {Influence of the exchange screening
  parameter on the performance of screened hybrid functionals},\ }\href
  {https://doi.org/10.1063/1.2404663} {\bibfield  {journal} {\bibinfo
  {journal} {The Journal of Chemical Physics}\ }\textbf {\bibinfo {volume}
  {125}},\ \bibinfo {pages} {224106} (\bibinfo {year} {2006})}\BibitemShut
  {NoStop}%
\bibitem [{\citenamefont {Togo}\ \emph {et~al.}(2023)\citenamefont {Togo},
  \citenamefont {Chaput}, \citenamefont {Tadano},\ and\ \citenamefont
  {Tanaka}}]{phonopy1}%
  \BibitemOpen
  \bibfield  {author} {\bibinfo {author} {\bibfnamefont {A.}~\bibnamefont
  {Togo}}, \bibinfo {author} {\bibfnamefont {L.}~\bibnamefont {Chaput}},
  \bibinfo {author} {\bibfnamefont {T.}~\bibnamefont {Tadano}},\ and\ \bibinfo
  {author} {\bibfnamefont {I.}~\bibnamefont {Tanaka}},\ }\bibfield  {title}
  {\bibinfo {title} {Implementation strategies in phonopy and phono3py},\
  }\href {https://doi.org/10.1088/1361-648X/acd831} {\bibfield  {journal}
  {\bibinfo  {journal} {J. Phys. Condens. Matter}\ }\textbf {\bibinfo {volume}
  {35}},\ \bibinfo {pages} {353001} (\bibinfo {year} {2023})}\BibitemShut
  {NoStop}%
\bibitem [{\citenamefont {Togo}(2023)}]{phonopy2}%
  \BibitemOpen
  \bibfield  {author} {\bibinfo {author} {\bibfnamefont {A.}~\bibnamefont
  {Togo}},\ }\bibfield  {title} {\bibinfo {title} {First-principles phonon
  calculations with phonopy and phono3py},\ }\href
  {https://doi.org/10.7566/JPSJ.92.012001} {\bibfield  {journal} {\bibinfo
  {journal} {J. Phys. Soc. Jpn.}\ }\textbf {\bibinfo {volume} {92}},\ \bibinfo
  {pages} {012001} (\bibinfo {year} {2023})}\BibitemShut {NoStop}%
\bibitem [{\citenamefont {Hellman}\ \emph {et~al.}(2011)\citenamefont
  {Hellman}, \citenamefont {Abrikosov},\ and\ \citenamefont {Simak}}]{tdep1}%
  \BibitemOpen
  \bibfield  {author} {\bibinfo {author} {\bibfnamefont {O.}~\bibnamefont
  {Hellman}}, \bibinfo {author} {\bibfnamefont {I.~A.}\ \bibnamefont
  {Abrikosov}},\ and\ \bibinfo {author} {\bibfnamefont {S.~I.}\ \bibnamefont
  {Simak}},\ }\bibfield  {title} {\bibinfo {title} {Lattice dynamics of
  anharmonic solids from first principles},\ }\href
  {https://doi.org/10.1103/PhysRevB.84.180301} {\bibfield  {journal} {\bibinfo
  {journal} {Phys. Rev. B}\ }\textbf {\bibinfo {volume} {84}},\ \bibinfo
  {pages} {180301} (\bibinfo {year} {2011})}\BibitemShut {NoStop}%
\bibitem [{\citenamefont {Hellman}\ and\ \citenamefont
  {Abrikosov}(2013)}]{tdep2}%
  \BibitemOpen
  \bibfield  {author} {\bibinfo {author} {\bibfnamefont {O.}~\bibnamefont
  {Hellman}}\ and\ \bibinfo {author} {\bibfnamefont {I.~A.}\ \bibnamefont
  {Abrikosov}},\ }\bibfield  {title} {\bibinfo {title} {Temperature-dependent
  effective third-order interatomic force constants from first principles},\
  }\href {https://doi.org/10.1103/PhysRevB.88.144301} {\bibfield  {journal}
  {\bibinfo  {journal} {Phys. Rev. B}\ }\textbf {\bibinfo {volume} {88}},\
  \bibinfo {pages} {144301} (\bibinfo {year} {2013})}\BibitemShut {NoStop}%
\bibitem [{\citenamefont {Shulumba}\ \emph {et~al.}(2017)\citenamefont
  {Shulumba}, \citenamefont {Hellman},\ and\ \citenamefont {Minnich}}]{stdep}%
  \BibitemOpen
  \bibfield  {author} {\bibinfo {author} {\bibfnamefont {N.}~\bibnamefont
  {Shulumba}}, \bibinfo {author} {\bibfnamefont {O.}~\bibnamefont {Hellman}},\
  and\ \bibinfo {author} {\bibfnamefont {A.~J.}\ \bibnamefont {Minnich}},\
  }\bibfield  {title} {\bibinfo {title} {Intrinsic localized mode and low
  thermal conductivity of pbse},\ }\href
  {https://doi.org/10.1103/PhysRevB.95.014302} {\bibfield  {journal} {\bibinfo
  {journal} {Phys. Rev. B}\ }\textbf {\bibinfo {volume} {95}},\ \bibinfo
  {pages} {014302} (\bibinfo {year} {2017})}\BibitemShut {NoStop}%
\bibitem [{\citenamefont {Krause}\ \emph {et~al.}(2018)\citenamefont {Krause},
  \citenamefont {Kuznetsov}, \citenamefont {Yakshin}, \citenamefont
  {Ibrahimkutty}, \citenamefont {Baumbach},\ and\ \citenamefont
  {Bijkerk}}]{Krause_nb5222}%
  \BibitemOpen
  \bibfield  {author} {\bibinfo {author} {\bibfnamefont {B.}~\bibnamefont
  {Krause}}, \bibinfo {author} {\bibfnamefont {D.~S.}\ \bibnamefont
  {Kuznetsov}}, \bibinfo {author} {\bibfnamefont {A.~E.}\ \bibnamefont
  {Yakshin}}, \bibinfo {author} {\bibfnamefont {S.}~\bibnamefont
  {Ibrahimkutty}}, \bibinfo {author} {\bibfnamefont {T.}~\bibnamefont
  {Baumbach}},\ and\ \bibinfo {author} {\bibfnamefont {F.}~\bibnamefont
  {Bijkerk}},\ }\bibfield  {title} {\bibinfo {title} {{{\it In situ} and
  real-time monitoring of structure formation during non-reactive sputter
  deposition of lanthanum and reactive sputter deposition of lanthanum
  nitride}},\ }\href {https://doi.org/10.1107/S1600576718007367} {\bibfield
  {journal} {\bibinfo  {journal} {Journal of Applied Crystallography}\ }\textbf
  {\bibinfo {volume} {51}},\ \bibinfo {pages} {1013} (\bibinfo {year}
  {2018})}\BibitemShut {NoStop}%
\end{thebibliography}%
\end{document}



\title{Supplemental Material for \textquotedblleft Uncovering Antipolar Ordering and and Pressure-Tunable Phases in Hexagonal LaN\textquotedblright}

\author{Atanu Paul}
\affiliation{Smart Ferroic Materials Center, Physics Department and Institute for Nanoscience and Engineering, University of Arkansas, Fayetteville, Arkansas 72701, USA}

\author{Laurent Bellaiche}
\affiliation{Smart Ferroic Materials Center, Physics Department and Institute for Nanoscience and Engineering, University of Arkansas, Fayetteville, Arkansas 72701, USA}
\affiliation{Department of Materials Science and Engineering, Tel Aviv University, Ramat Aviv, Tel Aviv 6997801, Israel}%

\author{Charles Paillard}
\affiliation{Smart Ferroic Materials Center, Physics Department and Institute for Nanoscience and Engineering, University of Arkansas, Fayetteville, Arkansas 72701, USA}
\affiliation{Université Paris-Saclay, CentraleSupélec, CNRS, Laboratoire SPMS, Gif-sur-Yvette, 91190 France.}

\date{\today}

\maketitle

\section{Method}
The first-principles density functional theory (DFT) calculations as implemented in the Vienna Ab initio Simulation Package (VASP)~\cite{VASP1,VASP2} have been carried out for this work. The electron-ion interactions were treated within the projector augmented-wave (PAW) formalism~\cite{PAW}, and the exchange-correlation effects were described using the Perdew-Burke-Ernzerhof (PBE) functional~\cite{PBE}. A plane-wave energy cutoff of 640 eV was employed to expand the Kohn-Sham wave functions. A $\Gamma$-centered $k$-point mesh of 10$\times$10$\times$8 was used for the Brillouin zone integration. The energy convergence criteria was set to $10^{-8}$ eV. Atomic positions were relaxed until the residual forces on each atom were smaller than $10^{-5}$ eV/\AA. In order to calculate the band gap, we considered the Heyd-Scuseria-Ernzerhof (HSE06) hybrid functional~\cite{hse1,hse2} with an Hartree-Fock exchange parameters, $\alpha_x$ of 0.25. Considering in mind the expensiveness of the calculation, we considered the energy cutoff of 500 eV and a $\Gamma$-centered $k$-point mesh of $8\times8\times6$.

Phonon dispersions within the harmonic approximation were calculated using the finite-displacement supercell method as implemented in the VASP and Phonopy packages~\cite{phonopy1,phonopy2}. Symmetry-allowed atomic displacements were generated in a $5\times5\times3$ supercell, and the corresponding forces were obtained from first-principles calculations. The resulting interatomic force constants were then used to construct and diagonalize the dynamical matrix to obtain the phonon dispersion. Finite-temperature phonon dispersions were further evaluated using the temperature-dependent effective potential (TDEP) method combined with an efficient stochastic phase-space sampling scheme~\cite{tdep1,tdep2,stdep}. Within the TDEP framework, an effective potential energy surface is constructed to describe finite-temperature lattice dynamics. The force constants of this effective model are extracted via a least-squares fit to forces obtained from first-principles calculations. Instead of generating thermally disordered configurations using \textit{ab-initio} molecular dynamics (AIMD), the stochastic phase-space sampling approach is employed to directly generate the thermally representative atomic configurations.

\begin{table}
\caption{\label{tab:table1}
Theoretically calculated lattice parameters, $a$, $c$, axial ratio ($c$/$a$), polar displacement (\AA) - displacement from each La layer from its nearby N-layer, relative energy (meV/f.u.), band gap (eV) using HSE06 and PBE, polarization ($\mu \text{C}/\text{cm}^2$) in each La-N sublattice, total polarization ($P$) ($\mu \text{C}/\text{cm}^2$) for the studied hexagonal structures (H5, WZ, AP) and experimental lattice parameters of the WZ phase. 
}

\centering   

\begin{ruledtabular}
\begin{tabular}{ccccc @{\hspace{0.3cm}\vrule width 0.5pt \hspace{0.0cm}} ccccccc}
Phase & \shortstack{Space \\ group} & Method & \shortstack{a \\ (\AA)} & \shortstack{c \\ (\AA)}  & c/a &
\shortstack{Polar \\ displa- \\cement \\ (\AA)} & \shortstack{Energy \\ (meV/f.u.)} & \shortstack{HSE06 \\ Band \\ gap \\ (eV)} & \shortstack{PBE \\ Band \\ gap \\ (eV)} & \shortstack{$P$ in each \\ sublattice \\ ($\mu\text{C}/\text{cm}^2$)} & \shortstack{Total \\ $P$ \\ ($\mu\text{C}/\text{cm}^2$)} \\
\hline
H5 & $P6_{3}/mmc$ & Theory & 4.30 & 5.39  & 1.25 & 0.00 & 68.58 & 1.50 & 0.43 & 0.00 & 0.00 \\
WZ & $P6_{3}mc$   & Theory & 4.13 & 5.99  & 1.45 & 0.53 &  0.00 & 2.23 & 1.17 & 33.50 & 67.00\\
   & $P6_{3}mc$   & Expt.~\cite{Krause_nb5222}  & 4.08 & 5.84  & 1.43 &      &       &      &       & \\
AP & $P\bar{3}m1$ & Theory & 4.28 & 5.46  & 1.28 & 0.12 & 62.11 & 1.56 & 0.74  &  9.00 & 0.00 \\
\end{tabular}
\end{ruledtabular}
\end{table}

\section{Structural and electronic properties}
The structural and electronic properties of the H5, AP and WZ phases are presented in Table \ref{tab:table1}.

\section{Enthalpy and volume under pressure of LaN}
\begin{figure}[t]
\includegraphics[width=14.0cm]{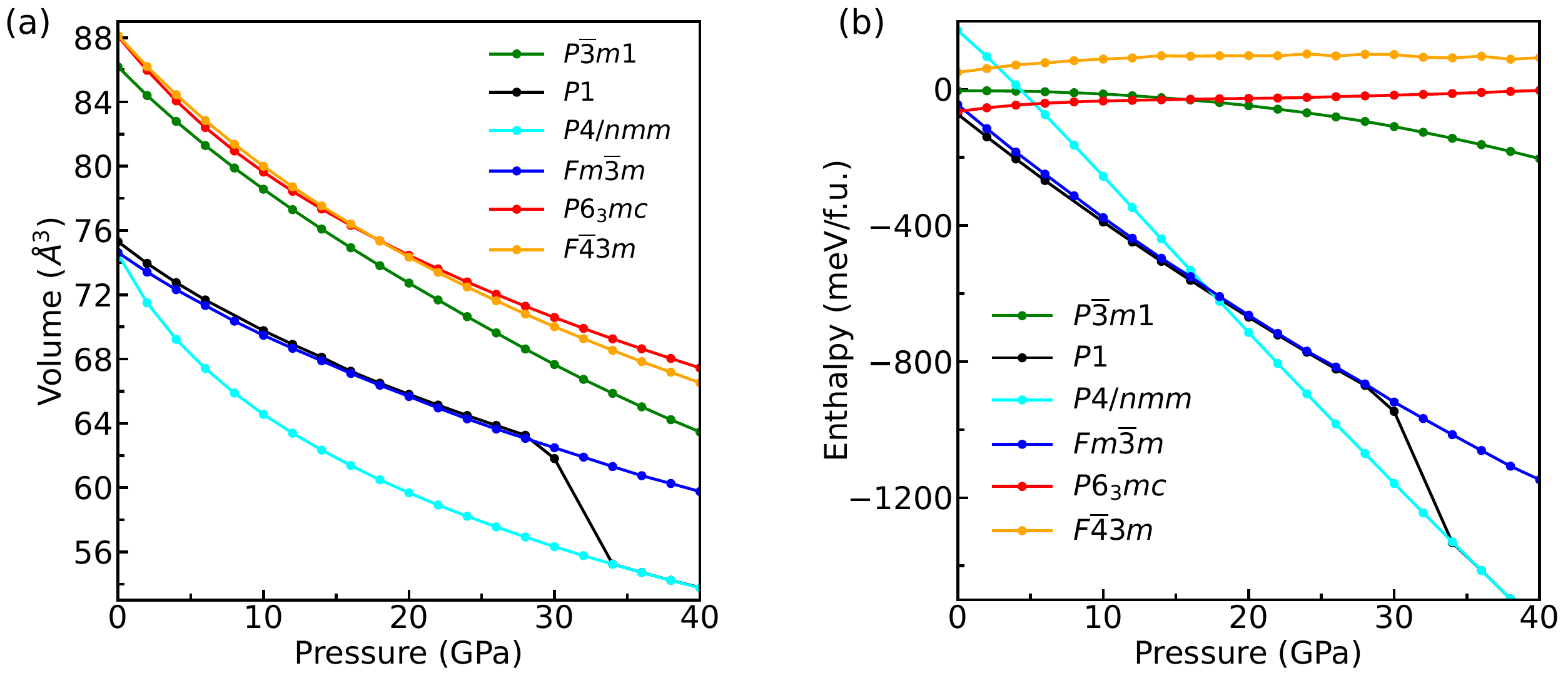}
\caption{\label{Fig1} (a) Volume (per two formula units of LaN) with applied pressure for different crystal symmetries ($P\bar{3}m1$ (antipolar), $P1$ (distorted rock-salt), $P4/nmm$ (tetragonal), $Fm\bar{3}m$ (rock-salt), $P6_{3}mc$ (wurtzite), and $F\bar{4}3m$ (zinc-blende)) of LaN. (b) Enthalpy with applied pressure for the different crystal symmetries. The zero of the enthalpy is set to that of the H5 phase of LaN.}
\end{figure}

To examine the effect of pressure on the considered phases of LaN, we calculated the volume and enthalpy as functions of pressure, as shown in Figs.~\ref{Fig1}(a) and (b), respectively. The volume of all phases decreases monotonically with increasing pressure. Notably, the volume of the distorted rock-salt structure ($P1$) approaches that of the tetragonal ($P4/nmm$) phase above $\sim$30 GPa, indicating a phase transition.

The enthalpy differences between the phases are small at low pressures, indicating that their relative stability is sensitive to external perturbations. The antiperovskite (AP) phase becomes energetically favorable above $\sim$15 GPa with respect to the WZ phase. 
With increasing pressure, the enthalpies of the $P1$, $Fm\bar{3}m$, and $P4/nmm$ phases decrease more rapidly compared to the other structures.


\section{Contribution of energy and PV in the enthalpy of the WZ and AP phase}
\begin{figure}[t]
\includegraphics[width=16.0cm]{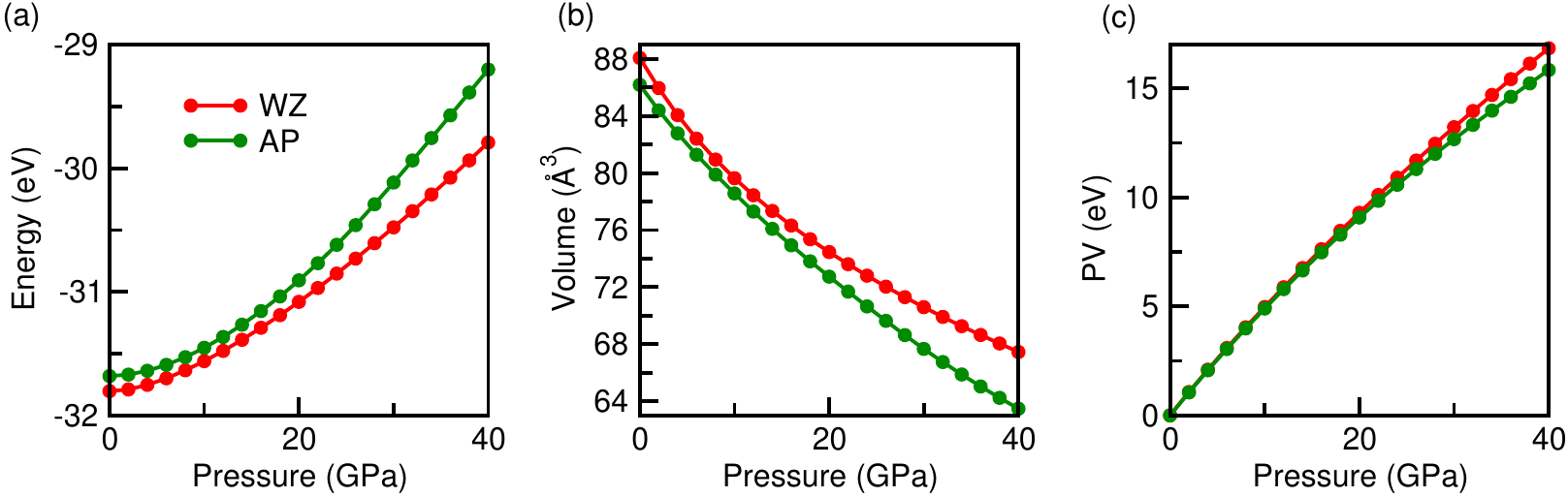}
\caption{\label{Fig_R1} Pressure ($P$) evolution of (a) energies, (b) volume ($V$), and (c) $PV$ for WZ and AP phases.}
\end{figure}

\begin{figure}[t]
\includegraphics[width=6.5cm]{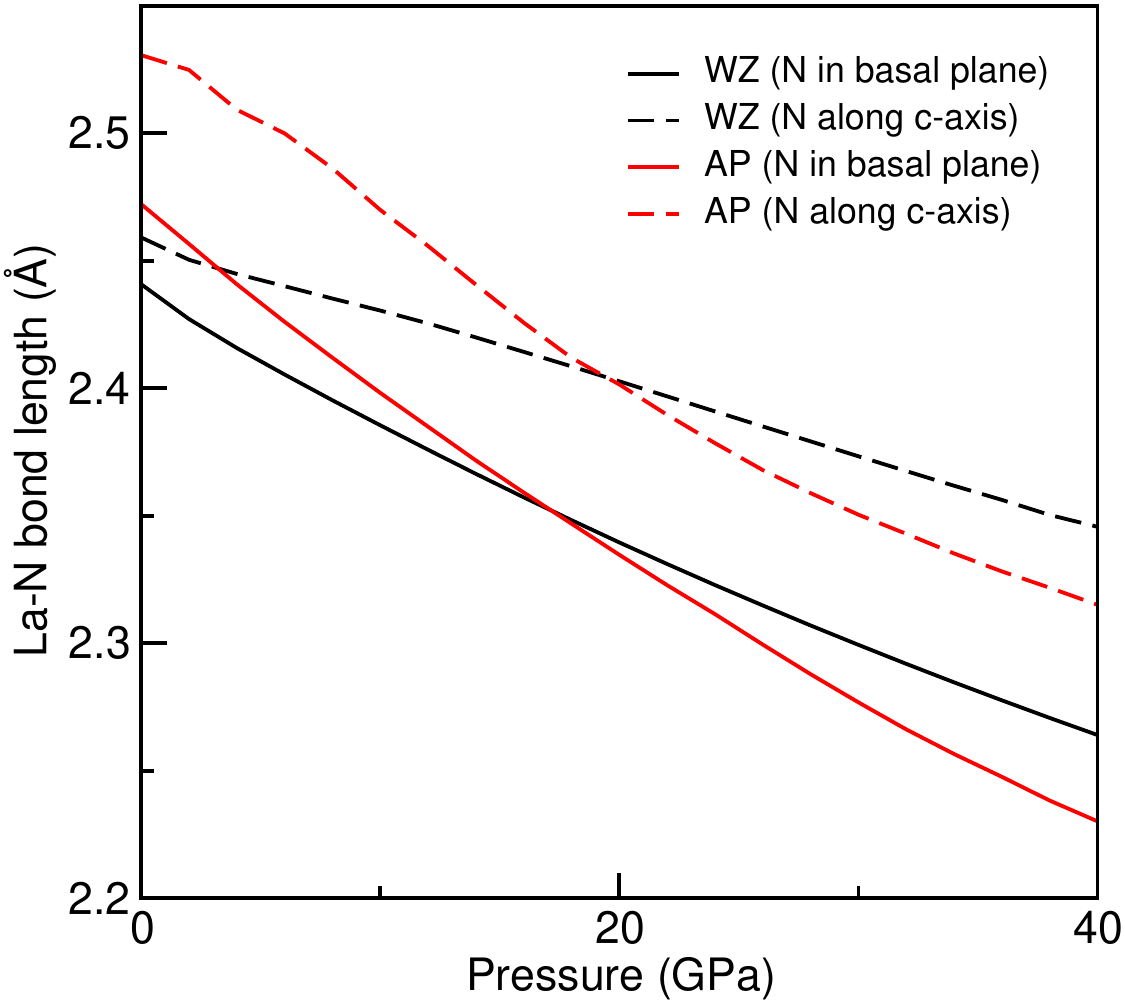}
\caption{\label{Fig_R2} Pressure evolution of the La-N bond lengths for the N in the basal plane and along $c$-axis.}
\end{figure}

To provide a deeper physical understanding of the different pressure evolution of stability of the WZ and AP phases, we decomposed the enthalpy ($H$ = $E$ + $PV$) into its internal energy ($E$) and pressure-volume ($PV$) contributions. The DFT-calculated internal energy increases with pressure for both phases, and the AP phase does not become energetically more favorable than the WZ phase (see Fig.~\ref{Fig_R1}  (a)). Therefore, the pressure-induced stabilization of the AP phase does not originate from the internal energy.

In contrast, the $PV$ contribution in the enthalpy increases with pressure for both phases but rises more rapidly for the WZ phase because it retains a larger volume under compression (Figs.~\ref{Fig_R1} (b) and (c)). Consequently, the enthalpy of the AP phase becomes lower than that of the WZ phase above approximately 15.5 GPa, demonstrating that the stabilization of the AP phase is primarily driven by its smaller volume.

To understand the origin of the smaller volume of the AP phase, we analyzed the pressure dependence of the La-N bond lengths, considering bonds in both the basal plane and along the $c$-axis (see Fig.~\ref{Fig_R2}). Although the La-N bond lengths decrease with pressure in both phases, the reduction is larger in the AP phase, indicating that the AP structure accommodates compression more efficiently through local bond shortening.

This distinct behavior is closely related to the different dipole arrangements in the two structures. In the WZ phase, all local dipoles are aligned parallel to one another. Under hydrostatic pressure, the increasing short-range repulsive interactions between neighboring ions favor a reduction of the polar off-centering, causing the polar distortion to decrease. As the cation and anion sublattices move closer to their centrosymmetric positions, the unit cell contracts less efficiently along the polar direction, resulting in a relatively larger equilibrium volume.

In contrast, the AP phase contains antiparallel neighboring dipoles whose macroscopic polarization cancels. Because the dipoles compensate each other, pressure does not primarily suppress the local off-centering. Instead, the structure accommodates compression by enhancing the antipolar displacements while simultaneously allowing greater La-N bond shortening. This cooperative rearrangement enables a more efficient packing of the ions, giving rise to a smaller unit-cell volume than in the WZ phase.

Consequently, the AP phase exhibits a lower $PV$ contribution under compression. Since the internal energy alone does not favor the AP phase, this volume advantage is responsible for the pressure-induced stabilization of the AP phase above approximately 15.5 GPa.

\section{Decomposition of the Energy Along the H5-WZ and AP-WZ Interpolation Paths}

\begin{figure}[t]
\includegraphics[width=16.0cm]{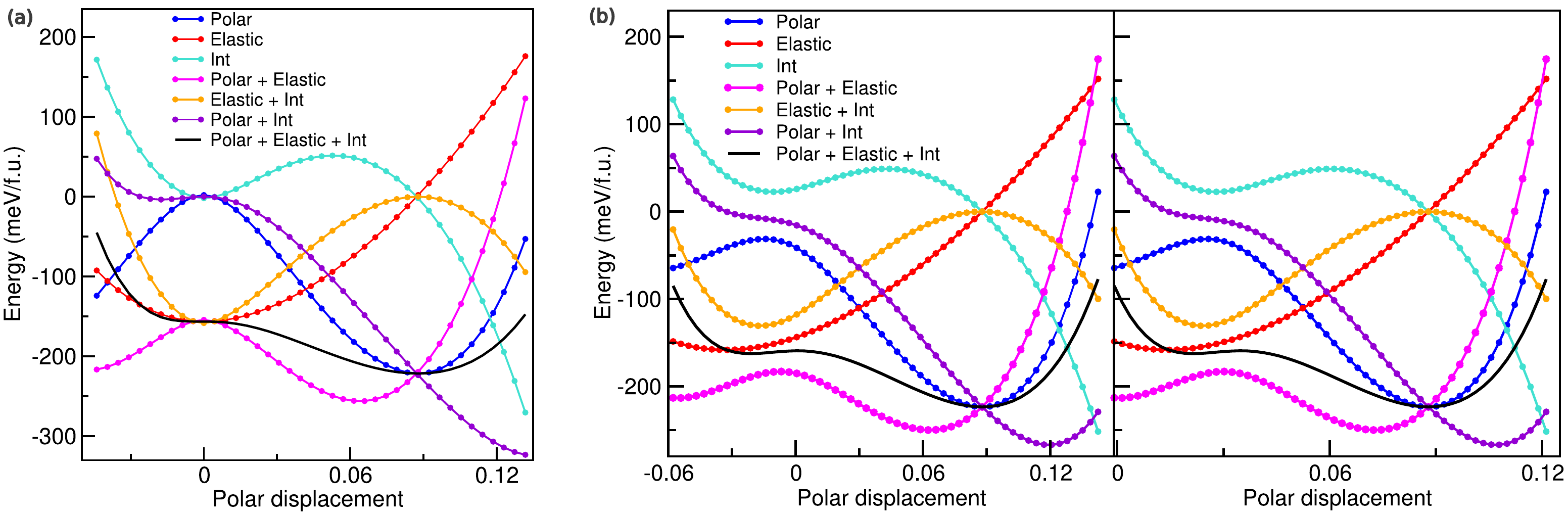}
\caption{\label{Fig_R3} Contributions from different degrees of freedom and their combined effect on the total energy as a function of polar displacement for the linearly interpolated structures along the (a) H5-WZ and (b) AP-WZ interpolation paths. The two panels in (b) correspond to the two polar displacement modes of the two sublattices.}
\end{figure}

The total energies ($E_{total}$) of the linearly interpolated structures along H5-WZ and AP-WZ can be approximately expressed as:
$$
E_{total} = E_{ref} + E_{polar} + E_{elastic} + E_{int}
$$

where, $E_{ref}$ is the energy of the reference nonpolar structure, for this work, the lattice parameters of the reference structure are chosen to that of the WZ phase. $E_{polar}$ represents the energy due to the pure polar (or antipolar) displacements from the reference nonpolar structure. $E_{elastic}$ represents the energy change associated with varying the axial ratio of the reference nonpolar structure, while keeping the atomic positions fixed at their corresponding nonpolar positions. $E_{int}$ represents the remaining energy arising due to the interaction between polar and elastic degree of freedom.

The contribution due to the polar (or antipolar) displacement, elastic deformation, polar and elastic interaction and their combined effect on the total energy ($E_{total}$) are shown in the Figs.~\ref{Fig_R3} for both cases, (a) H5 to WZ and (b) AP to WZ. In the figure we have plotted the energy contribution (meV/f.u.) with respect to the energy of the reference structure as a function of polar displacement. The polar (or antipolar) displacement and the (c/a) ratio used in this analysis are chosen to be exactly the same as those used in Fig. 3 of the manuscript.

For both pathways, the polar contribution exhibits a minimum close to the equilibrium polar distortion of the WZ phase as expected, and a maximum near the nonpolar (Fig. (a)) and antipolar (Fig. (b)) displacement. In contrast, the elastic contribution displays an approximately asymmetric parabolic behavior, with a minimum near the equilibrium axial ratio of the corresponding nonpolar (H5) or antipolar (AP) structure. Considering only the ionic contribution significantly overestimates the energy difference between the H5/AP  and the WZ structures. Inclusion of the elastic contribution substantially reduces this energy difference while preserving the minimum near the WZ polar displacement. Finally, incorporating the interaction energy between the polar and elastic degrees of freedom together with the individual polar and elastic contributions reproduces the total energy profile along the interpolated pathways. The main variation of the switching energy landscape originates from the polar degrees of freedom, while changes in the axial ratio provide an important contribution to modify the energy landscape.

\section{Born effective charges under pressure of LaN}
\begin{figure}[t]
\includegraphics[width=13.0cm]{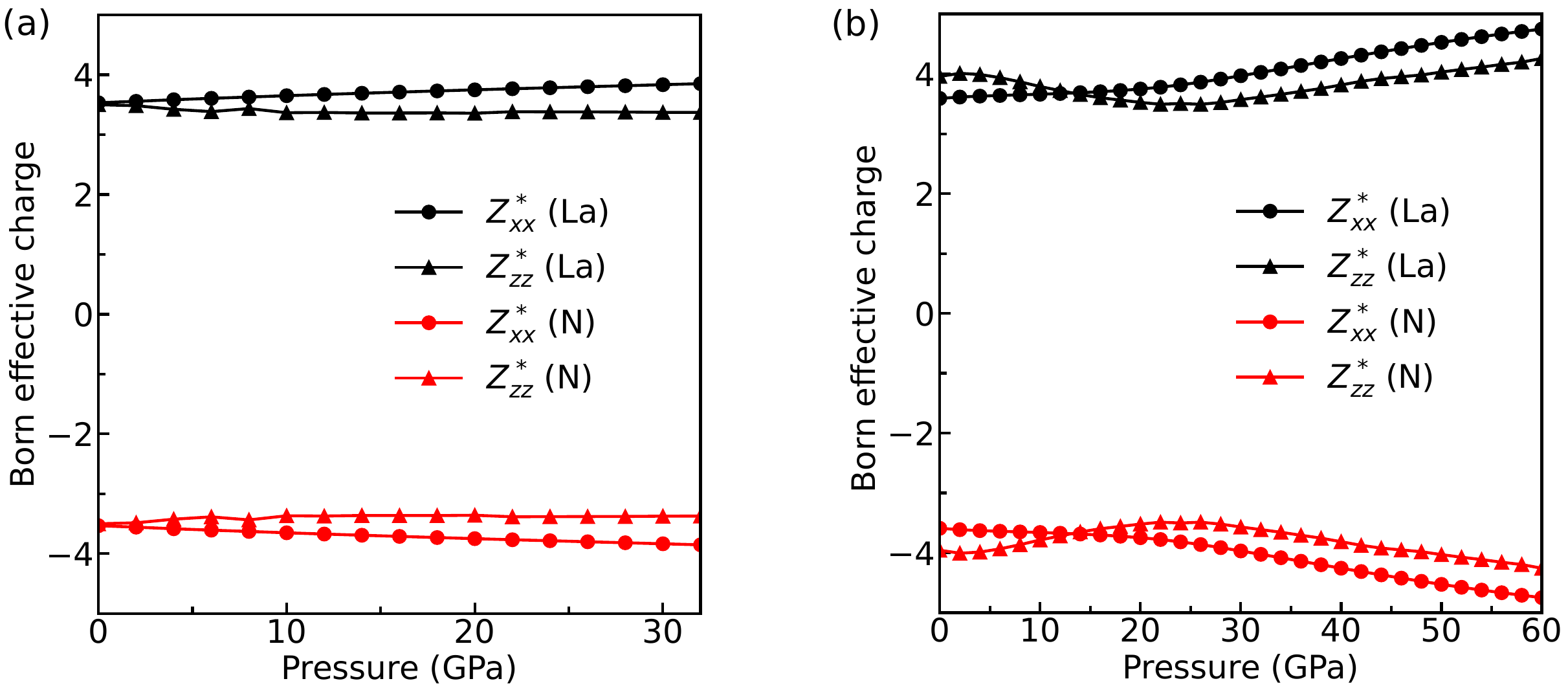}
\caption{\label{Fig2} Born effective charges under pressure for LaN in the (a) wurtzite and (b) antipolar phases.}
\end{figure}

\begin{figure}[t]
\includegraphics[width=7.0cm]{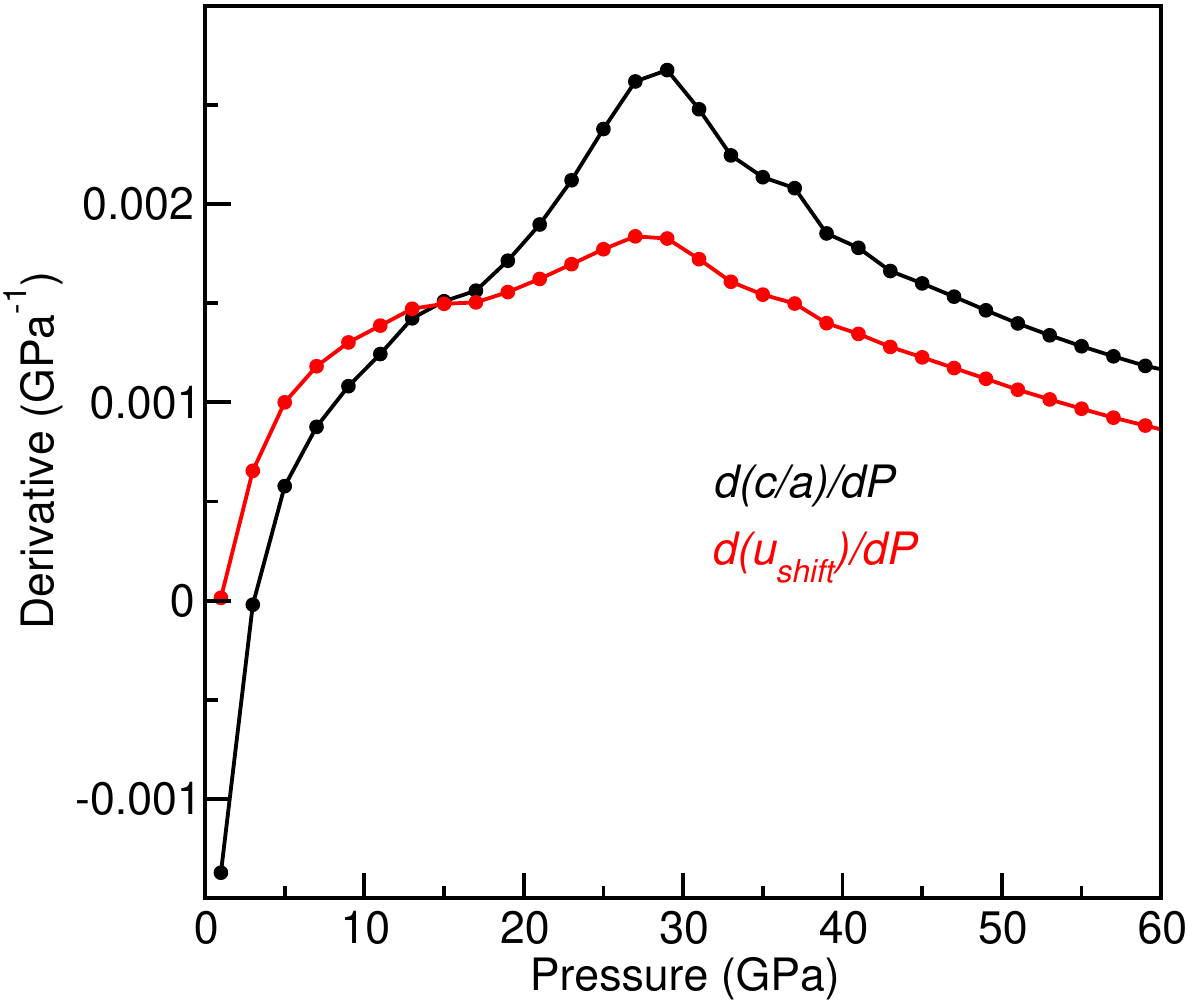}
\caption{\label{Fig2_2} Pressure dependence of the derivatives of the axial ratio ($c/a$) and polar displacement ($u_{\mathrm{shift}}$) in the AP phase.}
\end{figure}

The Born effective charges of La and N under pressure are shown in Figs.~\ref{Fig2}(a) and (b) for the wurtzite and antipolar (AP) phases of LaN, respectively. These Born effective charges were used to calculate the polarization of each LaN$_{4}$ sublattice in the corresponding phases. Notably, the $Z^{*}_{zz}$ value of La in the AP phase [Fig.~\ref{Fig2}(b)] decreases with pressure up to $\sim$25 GPa and subsequently increases with further compression. This unusual behavior is related to the pressure dependence of the axial ratio and the polar displacement in the AP phase. To illustrate this correlation, the pressure derivatives of the axial ratio and the polar distortion are plotted as functions of pressure in Fig.~\ref{Fig2_2}. The variation of $Z^{*}_{zz}$ of La in Fig.~\ref{Fig2}(b) closely follows the behavior shown in Fig.~\ref{Fig2_2}, indicating a strong correlation between the Born effective charge and the pressure evolution of the structural parameters.


\section{Energy surface and saddle nature of the H5 phase}

\begin{figure}[t]
\includegraphics[width=16.0cm]{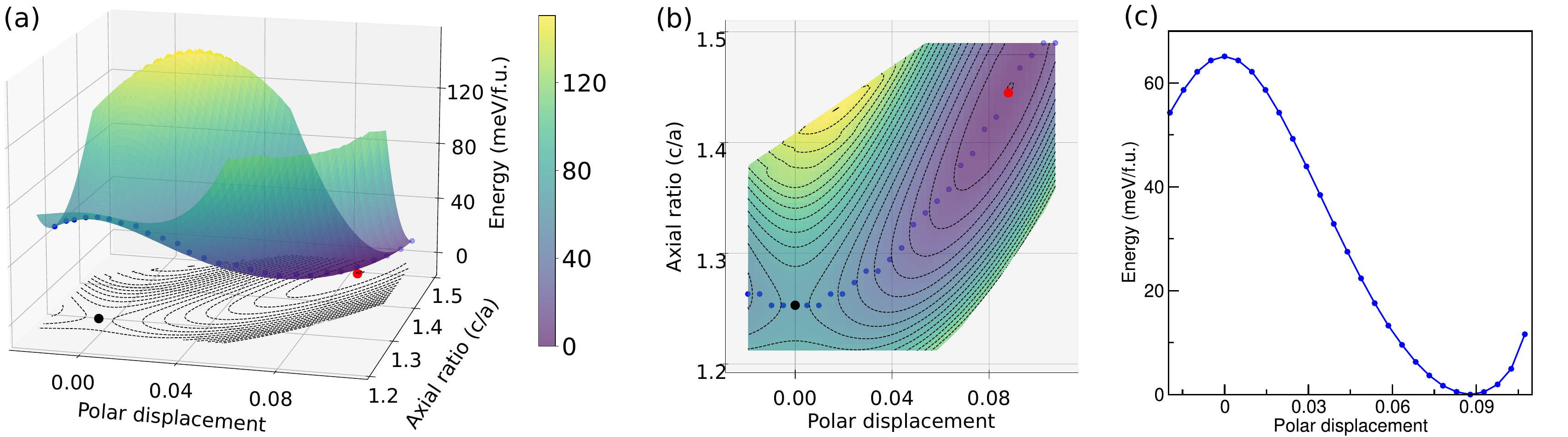}
\caption{\label{Fig3} (a) DFT-calculated energy surface as a function of axial ratio and polar displacement. The H5 and WZ phases are marked by black and red circles, respectively. The blue circles indicate the minimum-energy path connecting the WZ and H5 phases. (b) Top view of the energy surface (projection along the energy axis). (c) Energy profile along the minimum-energy path (blue circles) plotted as a function of polar displacement.}

\end{figure}

To elucidate the saddle-point nature of the H5 phase, we plot the DFT-calculated energy surface of LaN as a function of the axial ratio and polar displacement, encompassing both the H5 and WZ phases (see Fig.~\ref{Fig3}). The WZ phase is clearly located at a minimum of the energy surface, occurring at a polar displacement of 0.09 and an axial ratio of 1.45. The minimum energy path passes through the H5 phase, which is situated at a saddle point on the energy surface.

\section{Phonon dispersion of AP phase with supercell size}
\begin{figure}[t]
\includegraphics[width=8.0cm]{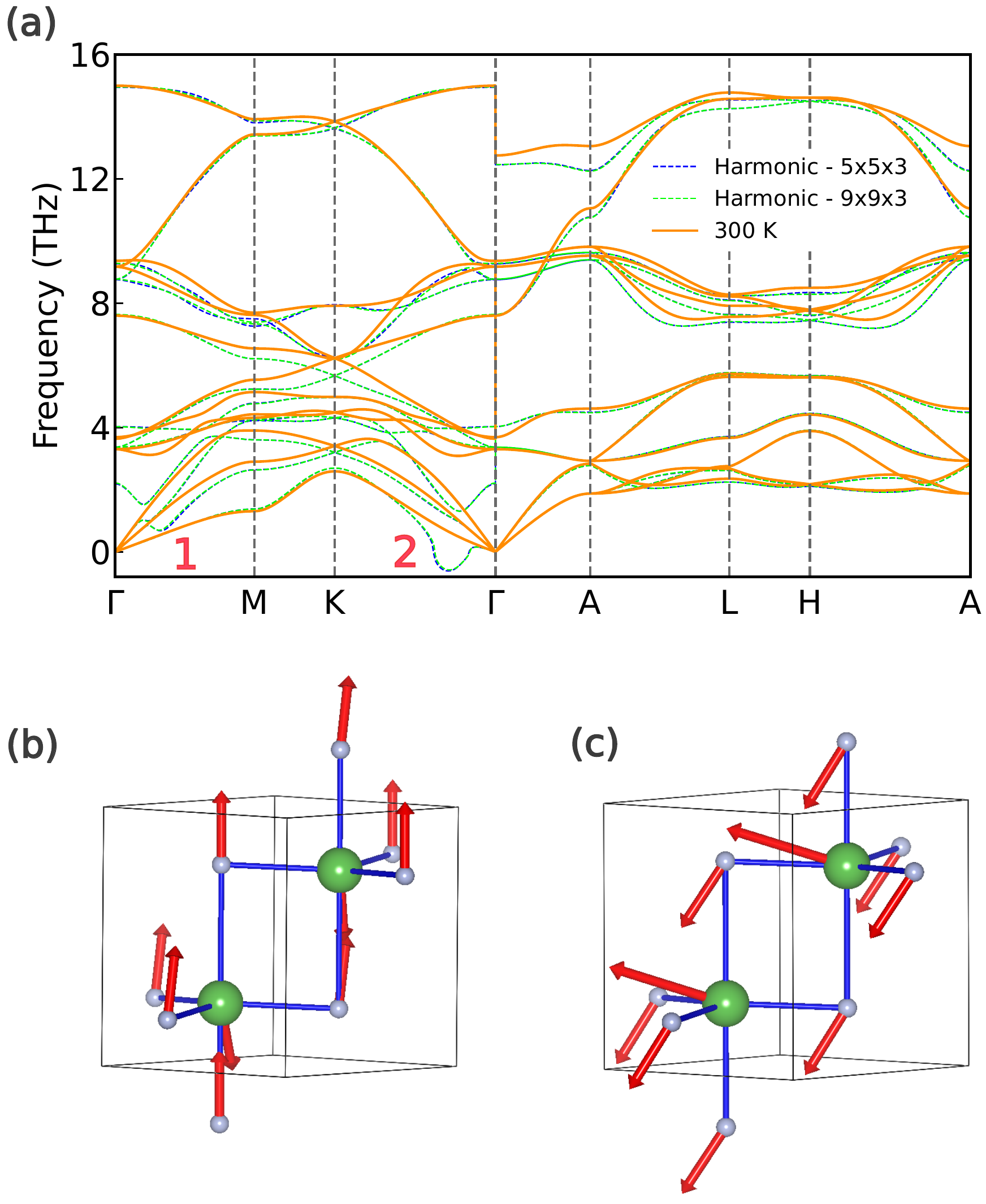}
\caption{\label{Fig8} LaN-AP: (a) Phonon dispersion of the AP phase calculated within the harmonic approximation considering supercell size of 5 $\times$ 5 $\times$ 3 and 9 $\times$ 9 $\times$ 3, and at 300 K considering supercell size of 5 $\times$ 5 $\times$ 3 in the hexagonal Brillouin zone. (b) and (c) correspond to the atomic displacements of the modes marked by 1 and 2, respectively, in the harmonic phonon dispersion.}

\end{figure}

The harmonic phonon dispersion of the AP phase was calculated using the supercell method with a (5 $\times$ 5 $\times$ 3) supercell. Upon reexamining the phonon dispersion, we find small positive frequency along the  $\Gamma$ - M (marked as 1) and small negative frequency along K - $\Gamma$ (marked as 2) directions, corresponding to wave vectors in the ($ab$)-plane of the hexagonal Brillouin zone, as shown in the Fig.~\ref{Fig8} (a). The atomic displacement patterns of these two modes are shown in Figs.~\ref{Fig8} (b) and (c). The small negative instability may lead to a potential polar wave at low temperature, which is not investigated here because of the large supercells required.

To further check the convergence of the phonon calculations, we repeated the calculations using a larger (9 $\times$ 9 $\times$ 3) supercell. The phonon dispersion changes very little, and the two modes remain.

\bibliography{LaN_AP_SM}